\documentclass[]{article}
\usepackage{times}
\usepackage{graphicx}
\usepackage{float}
\usepackage{caption}
\usepackage{epstopdf}
\usepackage{subcaption}
\usepackage[title]{appendix}
\usepackage{blindtext}
\usepackage{mathtools}
\usepackage{multicol}
\usepackage{lipsum}
\usepackage{amsmath}
\usepackage{authblk}
\usepackage{comment}
\usepackage{color}
\usepackage[scaled]{helvet}
\usepackage{cite}
\usepackage[]{authblk}
\usepackage[a4paper, portrait, margin=1in]{geometry}
\title{\bfseries Multi-phase-field modeling of microstructure evolution in metallic foams}
\author[a,b]{Samad Vakili\thanks{samad.vakili@rub.de}}
\author[a]{Ingo Steinbach\thanks{ingo.steinbach@rub.de}}
\author[a]{Fathollah Varnik\thanks{fathollah.varnik@rub.de \hspace*{0mm} (Corresponding author)}}
\affil[a]{\small Ruhr-Universit\"at Bochum, Interdisciplinary Center for Advanced Materials Simulation (ICAMS), Universit\"atsstr. 150, 44801 Bochum, Germany}
\date{}
\affil[b´]{\small Current Address: Max-Planck-Institut f\"ur Eisenforschung GmbH, Max-Planck-Stra\ss e 1, 40237 D\"usseldorf, Germany}

\renewcommand{\vec}[1]{\ensuremath{\mathbf{#1}}}

\newcommand{\sigmaGG}{\ensuremath{\sigma_\text{GG}}}
\newcommand{\sigmaLG}{\ensuremath{\sigma_\text{LG}}}
\newcommand{\rhoG}{\ensuremath{\rho_\text{Gas}}}
\newcommand{\rhoL}{\ensuremath{\rho_\text{Liq}}}
\newcommand{\cs}{\ensuremath{c_\text{s}}}
\newcommand{\dx}{\ensuremath{\Delta x}}
\newcommand{\dt}{\ensuremath{\Delta t}}
\newcommand{\Pid}{\ensuremath{\Pi_\text{disj}}}
\newcommand{\dpfilm}{\Delta p^{\text{film}}}
\begin{document}
\maketitle
\begin{abstract}
\setlength{\parindent}{0pt}
This paper represents a model for microstructure formation in metallic foams based on the multi-phase-field (MPF) approach. By the use of a no-coalescence boundary condition within this MPF-framework, it is possible to completely prevent coalescence of bubbles and thus focus on the formation of a closed porous microstructure. 
A modification of this non-wetting criterion allows for the controlled initiation of coalescence and the evolution of open structures.
The method is validated and used to simulate foam structure formation both in two and three dimensions.\\

\end{abstract}       

\section{Introduction}
\label{Introduction}

Metallic foams have a wide range of applications including shock absorbers, heat exchangers, load bearings, and catalysts \cite{Banhart2001}. Some of the important properties of these materials are high energy absorption, high compressibility, and bending stiffness \cite{Koerner2008,Yu1998, Gibson2000}. These properties are determined by the metallic character of the constituents, on the one hand, and the underlying porous microstructure, on the other hand. However, despite growing interest in using metallic foams, little is known about the mechanisms of microstructure formation in this class of materials.

This problem is at least partially related to the complexity of the processes involved in manufacturing metallic foams. A way to create bubbles is by subjecting the melt pool to a jet of gas~\cite{Leitlmeier2002}. Another, more frequent, route is to create a homogeneous distribution of bubbles in a melt pool via a suspension of solid particles, which start to produce gas above a threshold temperature~\cite{Akiyama1986}. Interestingly, the suspension mechanism is also used to producing porous structures in the polymer and the metal industry~\cite{Zhou2011, Koerner2008Art}. In this approach, the generated tiny bubbles grow by increasing the gas concentration until they form a space-filling porous microstructure, which is then frozen via solidification \cite{Banhart2006, Koerner2008, Lefebvre2008}. Since bubbles come in close contact during the growth process, a primary challenge is controlling the bubble coalescence, which usually proceeds significantly faster than bubble rearrangement dynamics. If not properly tuned, the coalescence of bubbles can lead to the formation of an open pore network with unfavorably large pores and a coarse microstructure.

A common approach to slow down the coalescence process in numerical studies is the use of a so-called disjoining potential~\cite{Saipavankumarbhogireddy2015} or pressure~\cite{deGennes2002}, which tends to keep bubbles apart~\cite{Banhart2006, Gergely2004, Wuebben2003, Lehmhus2010}. In its generic form, this is a reasonably simple approach to mimic the effective outcome of various physical forces, such as the one arising from the agglomeration of small particles between bubbles, Marangoni-type effects, interface elasticity, or a combination of these \cite{Babcsan2005, Koerner2002}.
To date, there are only a few models in the literature for studying microstructure formation in metallic foams \cite{Koerner2002, Koerner2008Art}. In these models, bubbles are simplified as empty spaces (voids), thus neglecting the  gas dynamics inside bubbles. 

In a recent study \cite{Vakili2020}, we proposed a simple approach to slow down the bubble-bubble coalescence process by tuning the interface energy alone. It was shown there that the rate of coalescence could be reduced by many orders of magnitude so that bubble rearrangements could occur prior to any coalescence. A disadvantage of the method was also mentioned, namely that the Laplace pressure, which scales linearly with the interface energy, was also reduced in a dramatic way, thus making the bubbles highly deformable \cite{Vakili2020}. In experiments, however, coalescence can be controlled without much effect of bubbles resistance to deformation. Here, we propose a qualitatively new approach that allows to completely prevent coalescence without any change of the liquid-gas interface energy. As a consequence, the bubbles' resistance to deformation remains unchanged. Moreover, a condition is also introduced to allow the initiation of coalescence if driving forces are strong enough to overcome a certain free energetic barrier.

The work in this paper is organized as follows. Section~\ref{sec:MPF} presents the new method and its important ingredients. The maturity of the proposed approach to simulate the formation of a complex foam microstructure is presented in section 3. A summary compiles the most important findings of this work.

\section{Simulation Method}
\label{sec:MPF}
The present simulation methodology combines ideas from multiphase flows~\cite{Brennen2005, Varnik2011, Moradi2011} with the well-established multi-phase-field method for microstructure evolution~\cite{Steinbach1996, Steinbach1999}. While the presence of fluid flow, capillarity, and wetting phenomena such as bubble-bubble coalescence and triple-phase contact angles in the formation of foams motivate the need for concepts in the field of multiphase flows, involving the multi-phase-field (MPF) method may, at least at first sight, appear less obvious. Therefore, we first introduce the MPF method and highlight the advantage of using such an approach. Dynamical equations, which dictate the time evolution of the system, are given after this subsection.

\subsection{A multi-phase-field model}
\label{subsec:Free-energy-functional}
\begin{figure}
	\centering
	\includegraphics[height=0.3\linewidth]{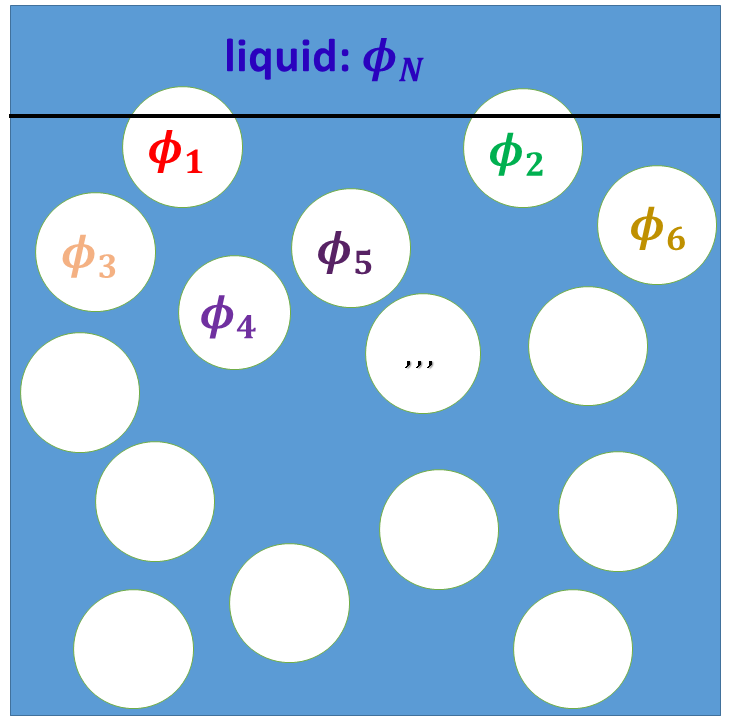}
	\includegraphics[width=0.4\linewidth]{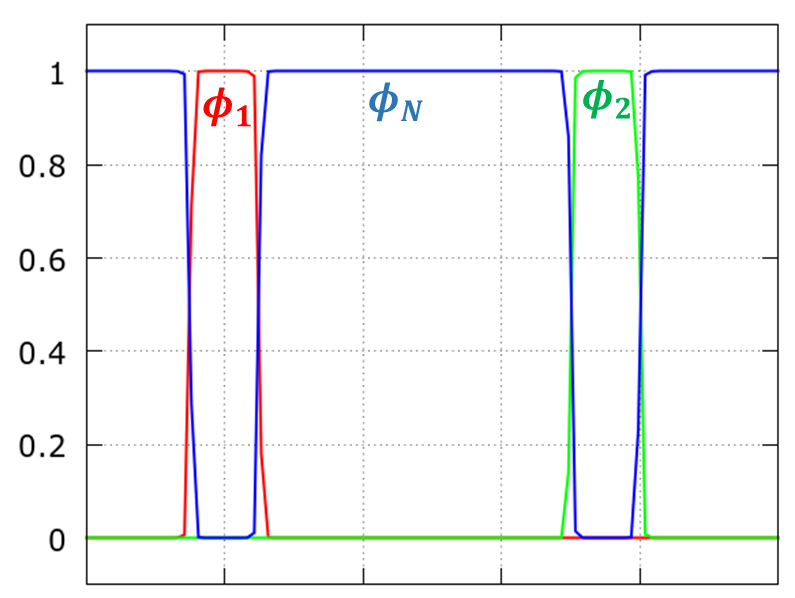}\\
	(a) \hspace{55mm} (b)\\
	\caption{(a) Schematic representation of the bubbles inside a melt. Each individual bubble and also the surrounding liquid is represented by a separate phase-field variable, $\phi_\alpha$, where $\alpha=1,...,N$. (b) The variation of phase field functions along the horizontal line drawn in (a).}
	\label{fig:multiphase_field}
\end{figure}

As mentioned above, the present approach builds upon the well-established multi-phase-field method. We use a version of this approach, which has first been proposed by Steinbach and coworkers \cite{Steinbach1996, Steinbach1999}. For this purpose, we consider each individual bubble as a separate 'phase' which occupies a certain region of space. The associated phase field function, $\phi_\alpha(\vec x)$, then takes the value of 1 if the point $\vec x$ is completely occupied by the bubble with index $\alpha$ ($\alpha=1,2,\ldots,N-1$; the index $N$ being reserved for the surrounding liquid). Similarly, $\phi_\alpha(\vec x)=0$ in the opposite case of complete absence of bubble $\alpha$ at point $\vec x$. Of more interest is, of course, the situation, where the point $\vec x$
lies at the interface of bubble $\alpha$ with other ones, in which case $0<\phi_\alpha(\vec x)<1$ (Fig.~\ref{fig:multiphase_field}). We adopt here the interpretation that $\phi_\alpha(\vec x)$ is the fraction of volume element $dV$ at $\vec x$ which is filled by the phase $\alpha$. From this, it immediately follows that $\sum_{\alpha=1}^N \phi_\alpha(\vec x)=1$, \label{page:sum-phi-is-one}where the index $\alpha=N$ accounts for the presence of ambient liquid.

While spatial distribution of bubbles is monitored by the phase field variables, physical properties of the system are accounted for in a free energy functional, $\cal F$. Following the standard multi-phase-field method, one writes~\cite{Steinbach1999,Boettinger2002}
\begin{equation}
\begin{aligned}
\mathcal{F} &= \int_{\Omega}^{} \mathcal{L}\big( \{ \phi \}, \{ \nabla \phi \} \big) dV,
\end{aligned}
\label{eq:F}
\end{equation}
where $\Omega$ denotes the integration domain, $\mathcal{L}$ is the free energy density, $\{ \phi \} = (\phi_{1},\phi_{2},...,\phi_{N} )$ and $\{ \nabla \phi \} = (\nabla \phi_{1},\nabla \phi_{2},..., \nabla \phi_{N})$.

One of the main purposes of setting up a free energy functional in the present study is that it allows to derive, in a systematic way, an expression for the pressure tensor. As will be seen below, the divergence of pressure tensor, $\nabla\cdot\mathrm{P}$, plays a central role in updating the fluid velocity field. Exploring the translational invariance of $\cal L$ and accounting for the constraint that the sum of all phase fields is conserved at any point in space, one obtains (for a derivation see 
	Appendix-\ref{appSec:derivation of pressure tensor}) 
\begin{equation}
	\nabla \cdot \mathbf{P} = -\frac{1}{N}\sum_{\alpha=1}^{N}\sum_{\beta=1}^{N} \bigg\{ \frac{\delta \mathcal{F}}{\delta \phi_{\alpha}} - \frac{\delta \mathcal{F}}{\delta \phi_{\beta}} \bigg\} \nabla\phi_{\alpha}.
	\label{eq:divP}
\end{equation}
Since all forces and thus $\nabla \cdot \mathbf{P}$ vanish at equilibrium, the right hand side of Eq.~(\ref{eq:divP}) can be understood as the sum of forces, which arise due to deviations from equilibrium state. By considering changes of the functional integral, Eq.~(\ref{eq:F}), under small deviations from thermodynamic equilibrium, one obtains,
\begin{equation}
	\frac{\delta \mathcal{F}}{\delta \phi_\alpha} = \frac{\partial \mathcal{L}}{\partial \phi_\alpha} - \nabla \cdot \frac{\partial \mathcal{L}}{\partial \nabla \phi_\alpha}.
	\label{eq:dFdphi}
\end{equation}
Derivatives on the right hand side of Eq.~(\ref{eq:dFdphi}) are to be understood as partial derivatives in ordinary sense. Under equilibrium conditions, the left hand side of Eq.~(\ref{eq:dFdphi}) vanishes and one obtains the well-known  Euler-Lagrange equation.

As the next necessary step, the free energy density, $\cal L$, must be specified. A standard choice here is a square-gradient model as in the Ginzburg-Landau theory of phase transitions~\cite{Penrose1990}. When extended to the case of multiple phases, it can be written as~\cite{Steinbach1999,Vakili2020,Schiedung2017} (see also Appendix-\ref{appSec:derivation of pressure tensor}),
\begin{equation}
\mathcal{L} = \sum_{\alpha = 1}^{N-1} \sum_{\beta = \alpha+1}^{N} \Big( -\frac{4 \sigma_{\alpha\beta}\eta}{\pi^2} \nabla \phi_{\alpha} \cdot \nabla \phi_{\beta} +  \frac{4\sigma_{\alpha \beta}}{\eta} |\phi_{\alpha} \phi_{\beta}| - 
\big[h(\phi_{\alpha}) p_{\alpha}(\rho_{\alpha}) + h(\phi_{\beta}) p_{\beta}(\rho_{\beta})\big] 
\Big),
\label{eq:L}
\end{equation}
where $\sigma_{\alpha \beta}$ is the interface energy between the phases $\alpha$ and $\beta$ and $\eta$ denotes the width of the interface. $|\phi_\alpha \phi_\beta|$ is known as the double obstacle (DO) potential~\cite{Steinbach2013} and the square bracket gives the contribution to pressure arising from the pair of phases $\alpha$ and $\beta$, with $h$ being an interpolation function. It can be shown that $h$ does not affect the equilibrium state of a planar interface~\cite{Vakili2020}. For a sphere in static equilibrium with its surrounding medium, on the other hand, the function $h$ enters the force balance condition (see Appendix-\ref{appSec:sphere}).

It is important to note that the pressures $p_\alpha$ and $p_\beta$ in Eq.~(\ref{eq:L}) must be evaluated via the equation of states (EOS) corresponding to the phases $\alpha$ and $\beta$, respectively. In the present study, the same ideal gas EOS is used for all bubbles ($\alpha=1,2,\ldots,N-1$). For the surrounding liquid ($\alpha=N$), we use the well-known van der Waals equation of state,
\begin{equation}
p_\alpha = 
\begin{cases}
c^2_\text{s,G} \rho_\alpha &  \alpha<N: \textrm{ Gas phase },\\
\dfrac{a\rho_\alpha}{b-\rho_\alpha} - c\rho_\alpha & \alpha=N: \textrm{ Liquid phase},
\end{cases}
\label{eq:EOS}
\end{equation}
where $c_\text{s,G}$ is the sound speed in the gas phase and $a$, $b$, and $c$ are the thermodynamic constants, characterizing the van der Waals liquid. It is noteworthy that $a$ and $c$ have the dimension of velocity square but $b$ is a unreachable threshold density, at which fluid pressure diverges.

Density is determined via,
\begin{equation}
	\rho_\alpha(t)=\dfrac{m_\alpha(t)}{V_\alpha(t)}=\dfrac{m_\alpha(t)}{\int_\Omega \phi_\alpha(\vec x,t)dV},
	\label{eq:density}
\end{equation}
with $m_\alpha(t)$ being the total mass of the $\alpha$-th phase at time $t$. The use of density from Eq.~(\ref{eq:density})
in evaluating pressure is justified because we consider processes which are slow compared to the speed of sound. Spatial variations of density are thus assumed to homogenize instantaneously both within the bubbles and in the surrounding liquid. The fact that we do not need to survey the process of sound propagation allows the choice of a coarser grid and a larger discretization time step and thus provides a major enhancement of computational efficiency.

\subsection{Dynamical equations}
\label{sec:fluid-dynamics}
In the present model, microstructure evolution is governed by the dynamics of phase fields, $\phi_\alpha$, with $\alpha \in \{1,2,\ldots,N\}$. The rule to update the $\phi$-fields accounts on the one hand for the dynamics of fluid and on the other hand for thermodynamic driving forces. Following~\cite{Subhedar2015,Subhedar2020}, one writes,
\begin{equation}
	\frac{\partial \phi_\alpha(\vec x)}{\partial t} + \mathbf{u}(\vec x) \cdot \nabla \phi_\alpha(\vec x) = -\frac{1}{\tilde{N}} \sum_{\beta=1}^{\tilde{N}} M_{\alpha\beta}\Big( \frac{\delta \mathcal{F}}{\delta \phi_\alpha(\vec x)} -\frac{\delta \mathcal{F}}{\delta \phi_\beta(\vec x)} \Big),
	\label{eq:phi-dot}
\end{equation}
where $\vec u$ is the fluid velocity, $\delta \mathcal{F} /\delta \phi_{\alpha}$ stands for functional or variational derivative as given by Eq.~(\ref{eq:dFdphi}) and $M_{\alpha \beta}$ is the interface mobility. Importantly, $\tilde{N}$ is the number of phase fields present at the point $\vec x$ and must be distinguished from the total number of phase fields $N$.

The fluid velocity field, $\vec u(\vec x)$, is tracked everywhere in space by solving Navier-Stokes (NS) equations,
\begin{equation}
	\rho(\vec x)\Big( \frac{\partial \vec u}{\partial t} + \vec u \cdot \nabla \vec u \Big) = -\nabla \cdot \mathbf{P} + \nabla \cdot \Big(\mu(\vec x) \big[\nabla \vec u +(\nabla \vec u)^T \big] \Big) + \vec f^\mathrm{ext.},
	\label{Navier_Stokes}
\end{equation}
where $\mu$ is dynamic viscosity, $\rho$ is the fluid density and $\vec f^\mathrm{ext}$ is the external force per unit volume. In regions of space where only a single phase, say $\alpha$, is present, $\mu=\mu_\alpha$ and $\rho=\rho_\alpha$. In general, however, these quantities are defined via phase-averages, $\rho(\vec x)=\sum^N_{\alpha=1}\phi_\alpha(\vec x)\rho_\alpha$ and $\mu(\vec x)=\sum^N_{\alpha=1}\phi_\alpha(\vec x)\mu_\alpha$.
Even though it may not be apparent at first glance, Eq.~(\ref{Navier_Stokes}) also contains interface effects on the right hand side. This information is encoded in the divergence of pressure tensor, $\nabla \cdot \mathbf{P}$, which accounts both for the variation of hydrostatic pressure with density via the EOS and for interface forces by considering curvature terms, see Eq.~(\ref{eq:divP3}).

\subsection{Non-coalescing case: The contact angle}
\label{subsec:contact-angle}
\begin{figure}
	\centering
	\includegraphics[width=0.3\linewidth]{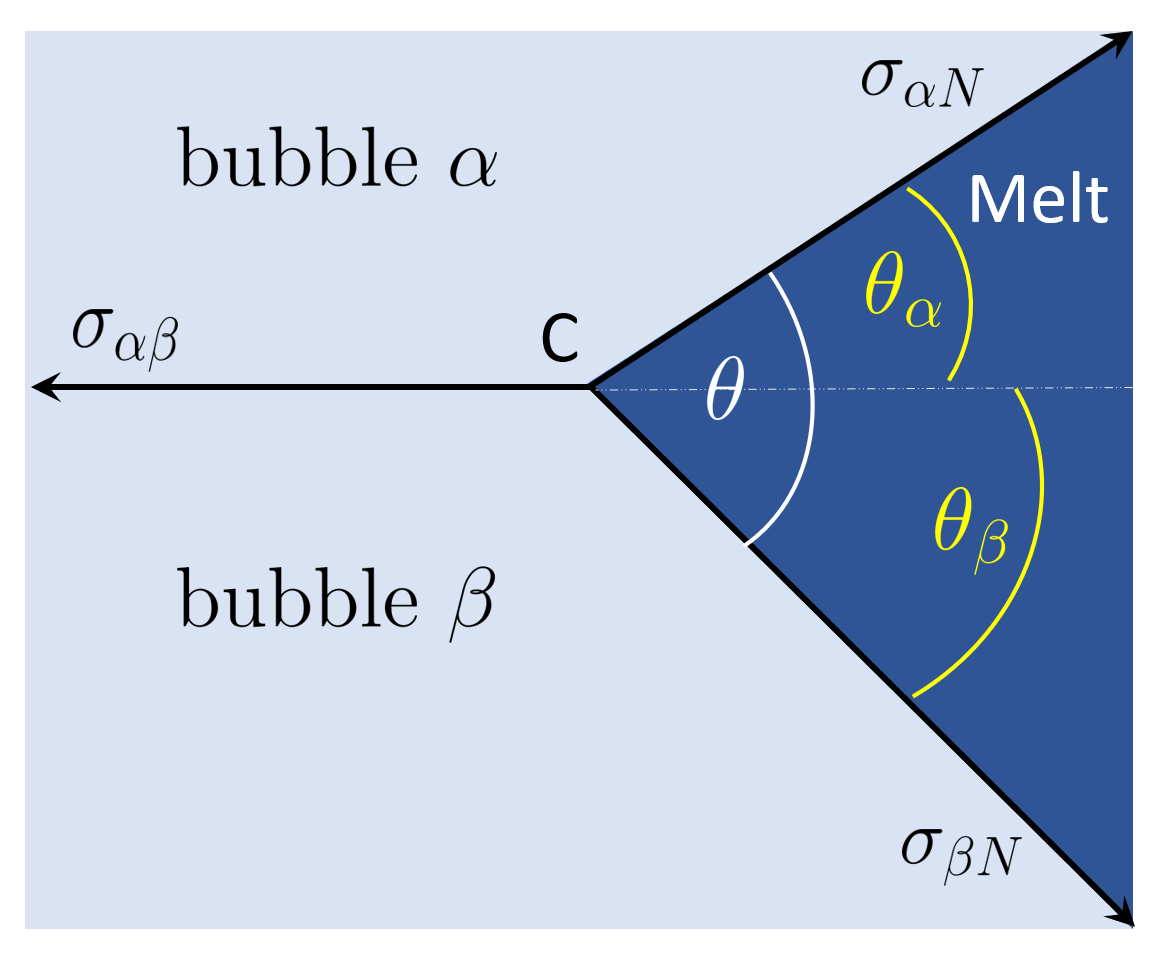}\hfill
	\includegraphics[width=0.29\linewidth]{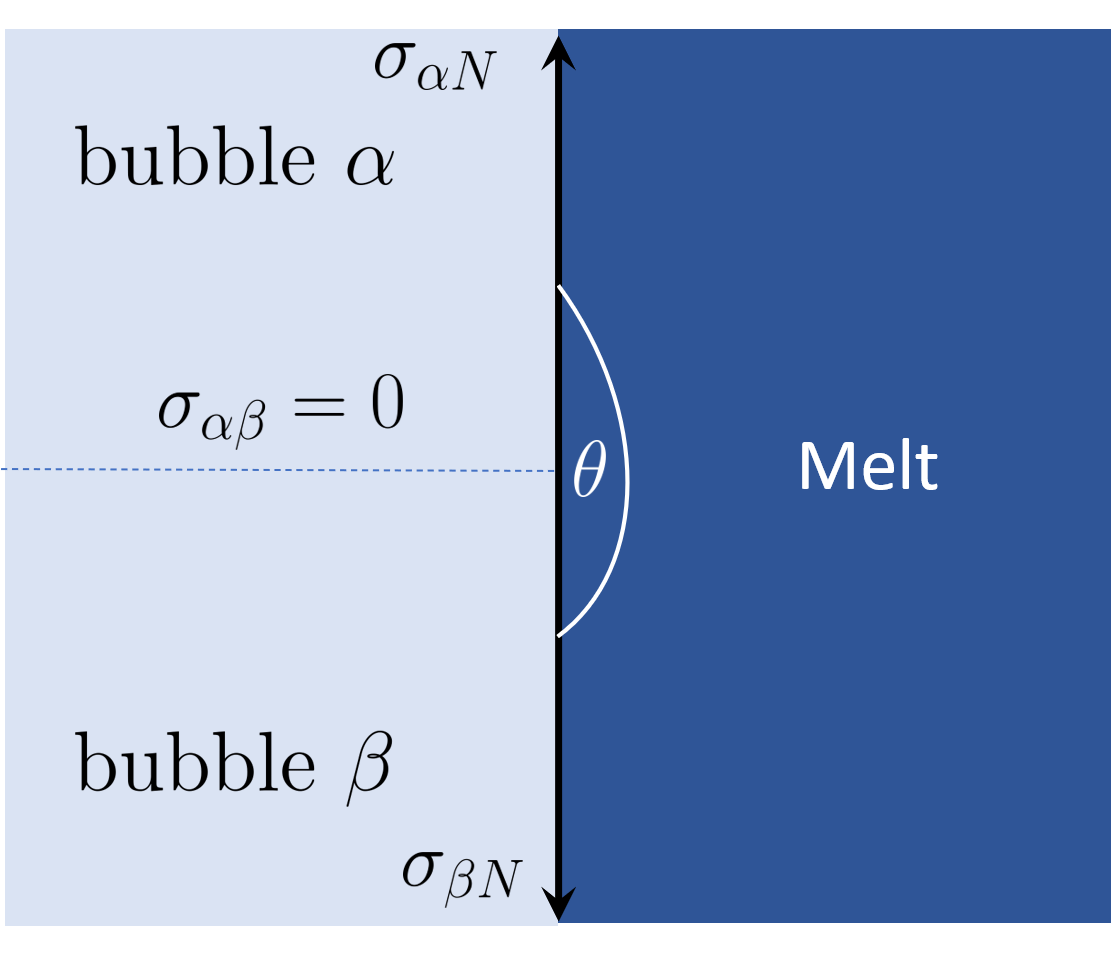}\hfill
	\includegraphics[width=0.3\linewidth]{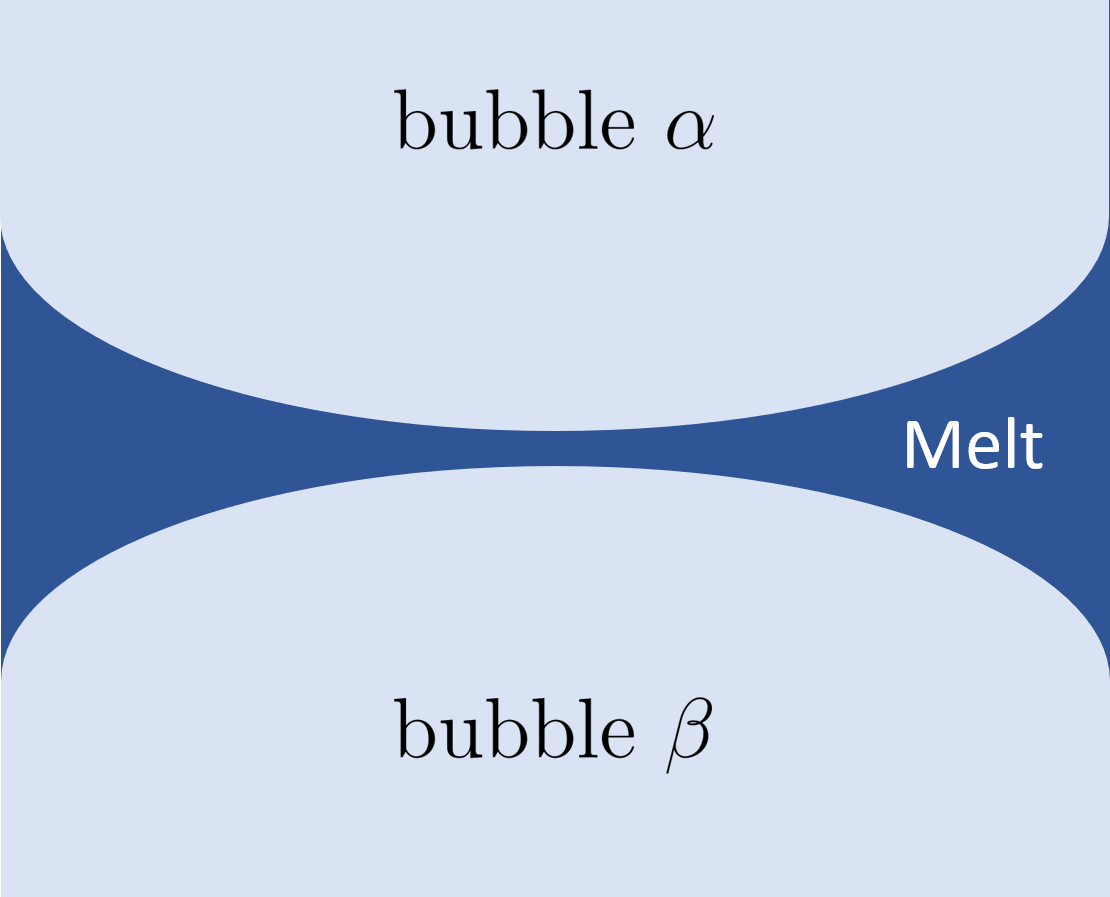}\\
	(a)\hspace*{0.33\textwidth}(b)\hspace*{0.33\textwidth} (c)
	\caption{A schematic view of two bubbles with different identities in contact, embedded in a melt pool. (a) At equilibrium, force balance at three phase contact line (the point C in this 2D projection) must be satisfied. When projected onto the $\sigma_{\alpha\beta}$-line, one obtains for the relation between specific surface free energies, $\sigma_{\alpha\beta}=\sigma_{\alpha N}\cos(\theta_\alpha)+\sigma_{\beta N}\cos(\theta_\beta)$. Balance of forces along the direction normal to the $\alpha\beta$-interface leads to $\sigma_{\alpha N}\sin(\theta_\alpha)=\sigma_{\beta N}\sin(\theta_\beta)$. In panel (b), the specific free energy for gas-gas contact vanishes ($\sigma_{\alpha\beta}=0$). It then follows from force balance that $\theta_\alpha=\pi/2$ and $\theta_\beta=\pi/2$. This means that the melt-gas and gas-gas interfaces must be perpendicular to each other. The horizontal dashed lines serves to highlight the fact that, within the multi-phase-field framework, bubbles $\alpha$ and $\beta$ remain distinguishable even though their interface energy vanishes. 
	Panel (c) shows what happens if $\sigma_{\alpha\beta}>\sigma_{\alpha N}+\sigma_{\beta N}$. In this case, bubbles go apart since the two newly formed liquid-gas interfaces have both together a lower free energy than the gas-gas interface.
	}
	\label{fig:contact-angle}
\end{figure}

The property that, within multi-phase-field method, one can assign a different phase index to each individual bubble provides a natural way to completely prevent the coalescence process. By doing so, each bubble can be treated as an independent entity with its physical properties. Importantly, there is also an interface energy parameter for each pair of bubbles, $\sigma_{\alpha\beta}$. By tuning this parameter, one can adjust the so-called contact-angle and thus control the overall structure of the foam.  This is illustrated in Fig.~\ref{fig:contact-angle} for the case of two bubbles in contact. At static equilibrium, it follows from the force balance that contact angles  $\theta_\alpha$ and  $\theta_\beta$ obey (Fig.~\ref{fig:contact-angle}a),
\begin{eqnarray}
\sigma_{\alpha N}\cos(\theta_\alpha) + \sigma_{\beta N}\cos(\theta_\beta)&=&\sigma_{\alpha\beta}\label{eq:contact-angle-par},\\
\sigma_{\alpha N}\sin(\theta_\alpha) - \sigma_{\beta N}\sin(\theta_\beta)&=&0.
\label{eq:contact-angle-per}
\end{eqnarray}

The above equations for contact angle deserve some comments. First, it can be easily verified that Eq.~(\ref{eq:contact-angle-par}) reduces to the well-known Young-equation~\cite{Young1805} for the three-phase contact angle on a flat solid. To see this, it is sufficient to identify the phase fields $\alpha$ and $\beta$ with vapor and solid phases, respectively, and use the fact that $\theta_\beta=0$ for a planar substrate. It is noteworthy that, in this case, Eq.~(\ref{eq:contact-angle-per}) is not valid since the normal projection of $\sigma_{\alpha N}$ is not compensated by that of $\sigma_{\beta N}$ but by elastic forces of the solid body, which resist deformation along the perpendicular direction. Second, Eq.~(\ref{eq:contact-angle-par}) makes sense only if $\sigma_{\alpha\beta}\le \sigma_{\alpha N}+\sigma_{\beta N}$ (recall that cosine function cannot exceed unity). This condition is always satisfied 
for $\sigma_{\alpha\beta}=0$. In this case, and recalling that all interface energies are non-negative, Eq.~(\ref{eq:contact-angle-par}) is solved by $\theta_\alpha = \theta_\beta = \pi/2$ (Fig.~\ref{fig:contact-angle}b). If, however, the gas-gas interface energy exceeds the sum of two gas-liquid interfaces energies, a melt layer forms between the bubbles, thereby reducing the total interface energy (Fig.~\ref{fig:contact-angle}c). This means that, from the perspective of bubble-bubble contact, a complete dewetting process takes place. The corresponding non-wetting condition can be expressed as 
\begin{equation}
\sigma_{\alpha\beta}=(\sigma_{\alpha N}+\sigma_{\beta N})(1+q), 
\label{eq:non-wetting}
\end{equation}
with $q$ being a non-negative empirical parameter. To see the physical meaning of $q$, let us consider that by some fluctuation an amount of liquid penetrates into a gas-gas interface and thus creates two liquid-gas interfaces (Fig.~\ref{fig:contact-angle}a$\to$c). If $q=0$, the total free energy will not change by this process, so that a fluctuation in the reverse direction can occur with equal probability and can restores the previous situation. A positive value of $q$ changes this balance in favor of dewetting. As will be shown below, it ensures the existence of a dewetting-force or, equivalently, a free energy barrier against bubble-bubble coalescence.

In the following, we will assume that all the bubbles have identical liquid-gas and gas-gas interface free energies, $\sigma_{\alpha N}=\sigma_{\beta N}\equiv\sigmaLG$ and $\sigma_{\alpha \beta}\equiv\sigmaGG$, respectively with $\alpha, \beta \in \{1,2,\ldots,N-1\}$. In this symmetric case, Eq.~(\ref{eq:contact-angle-per}) leads to $\theta_\alpha=\theta_\beta=\theta/2$ and Eq.~(\ref{eq:contact-angle-par}) becomes,
\begin{equation}
\cos\Big(\dfrac{\theta}{2}\Big) = \dfrac{\sigmaGG}{2\sigmaLG}.
\label{eq:contact-angle-sym}
\end{equation}
Equation~(\ref{eq:contact-angle-sym}) provides a simple benchmark problem to examine the proposed approach under static equilibrium condition.

A model which completely avoids coalescence is an idealization which allows to focus on a time window and parameter range, where bubbles rearrange and change their shapes in order to accommodate with boundary conditions, while at the same time keeping their individual character as a physical entity. This way, formation of closed structures containing a disconnected set of bubbles can be investigated. Such structures do usually have good mechanical properties, while at the same time being lighter than the corresponding bulk metallic solid.

\subsection{Condition for coalescence}
\label{subsec:coalescence-condition}
If one is interested in a study of open structures, a model that allows for coalescence is needed. An important application of open porous structures is heterogeneous catalysis, where  it is desirable that the reactants be able to enter and exit the porous body in order to come in contact with the entire catalytic surface. Other applications use open structures to reduce weight while at the same time optimizing mechanical properties.

Thanks to the flexibility of the multi-phase-field method, it is easy to include the coalescence phenomenon into the model, while at the same time having a safe control over its rate. This latter aspect is important since, as already mentioned above, the coalescence process usually proceeds quite fast, making it difficult to influence the foam structure by tuning the process parameters. 

Following a similar approach, which was used in a study of superalloys~\cite{Ali2020}, we introduce a simple criterion, 
for the initiation of coalescence. The basic idea is that two bubbles will coalesce if (i) they come sufficiently close so that their distance falls below a certain threshold and (ii) if the force pushing their respective liquid-gas interfaces towards each other, $\dpfilm$, is sufficiently high to overcome the free energy barriers (often referred to as disjoining pressure~\cite{deGennes2002}), which tends to keep the bubbles apart.

A natural choice for the threshold distance for coalescence is the interface width $\eta$. To estimate the disjoining pressure, which tries to push apart adjacent bubbles, we consider the slicing of a gas-gas interface into two adjacent liquid-gas interfaces. This process can be regarded as completed when the separation distance reaches the interface thickness, $\eta$. The change in free energy during this process is $dF=(\sigma_{\alpha N}+\sigma_{\beta N}-\sigma_{\alpha\beta})A=-q(\sigma_{\alpha N}+\sigma_{\beta N})A$, where we used Eq.~(\ref{eq:non-wetting}). $A$ is the interface area considered in this problem. The volume created during this process is $dV=A\eta$. The (disjoining) pressure, which is responsible for this process can now be obtained from the standard thermodynamic relation, $\Pid=-dF/dV$. This gives,
\begin{equation}
\Pid=q\,\dfrac{(\sigma_{\alpha N}+\sigma_{\beta N})}{\eta}.
\label{eq:Pidisj}
\end{equation}
It is important to note that the empirical parameter $q$ provides the possibility to freely adjust the disjoining pressure. Of course, in order for $\Pid$ to be effective, it must be of the same order of magnitude as the free energy densities involved in the problem. Thus, reasonable values of $q$ will be of the order of unity.

In order to obtain a closed expression for coalescence criterion, we estimate the driving force, $\dpfilm$, which pushes two neighboring gas-liquid interfaces, say $\alpha$ and $\beta$, towards each other. For this purpose, we first evaluate the net force per unit area on each of these interfaces. For the $\alpha$-th bubble, this force is given by 
$\Delta p_{\alpha N} = p_\alpha-(p_{N}+\kappa_\alpha\sigmaLG)$, where $p_\alpha$ is the inner bubble pressure, $p_N$ is that in the surrounding liquid and $\kappa_\alpha$ is the curvature of the gas-liquid interface on the side close to the $\alpha$-th bubble. Note that, as expected, this force vanishes at static equilibrium, where the Young-Laplace law holds.
This is in line with the idea that thermodynamic equilibrium is the state of matter where all forces are in balance (resulting in zero net force) and all macroscopic variables are time-independent.
Similarly, $\Delta p_{\beta N} = p_\beta-(p_{N}+\kappa_\beta\sigmaLG)$ is the net force per unit area on the $\beta$-th gas-liquid interface on the side facing the bubble $\alpha$. A careful analysis now reveals that the net force, which pushes the two bubbles towards each other is given by 
\begin{equation}
\dpfilm=\dfrac{\Delta p_{\alpha N}+\Delta p_{\beta N}}{2}.
\label{eq:dpfilm}
\end{equation}
Putting all this together, coalescence takes place within the present model if
\begin{equation}
	d<\eta \text{~~~and~~~} \dpfilm > \Pid \;\;\;\;(\text{coalescence-condition}),
	\label{eq:coalescence-condition}
\end{equation}
where $d$ is the distance between adjacent liquid-gas interfaces. If the condition (\ref{eq:coalescence-condition}) is satisfied, the region of space occupied by the two bubbles is identified as a single bubble. Technically, this is achieved by assigning the region of space occupied by $\phi_\beta$ to $\phi_\alpha$ (obviously, the reverse assignment $\alpha \to \beta$ is equally valid). The redundant phase-field (in this example $\phi_\beta$) is then removed from the list of phase fields. Moreover, since the interface associated with the contact area of the phases $\alpha$ and $\beta$ disappears, the corresponding specific free energy, $\sigma_{\alpha\beta}$, has no effect anymore and is thus removed from the list of parameters. Finally, depending on details of implementation, a re-indexing or other schemes can be applied to optimize memory usage.

This completes the model section. Next we will show that the model is indeed capable of both completely hindering or conditionally allowing the coalescence between neighboring bubbles (Fig.~\ref{fig:2Bubbles}). Then we will apply the thus developed computational tool to study evolution of microstructure in two and  three dimensions.

\section{Results and discussion}
\label{sec:Result}

As a necessary step, we perform a number of Benchmark simulations to validate the approach proposed above. Aiming to mimic the case of an open melt pool, the density (and consequently pressure) of the melt pool is kept constant. This corresponds to the situation that liquid can leave the simulation domain upon bubble growth. Interestingly, since the sum of all phase fields at any point in space is constrained to unity (see the first paragraph in section~\ref{subsec:Free-energy-functional} on page~\pageref{page:sum-phi-is-one}), the spatial domain occupied by $\phi_N$ (liquid phase) reduces automatically as the total gas volume increases. The insertion of mass into the bubbles is stopped when the volume occupied by the gas phase exceeds a certain threshold.

Unless otherwise stated, the grid spacing is $\dx = 0.01$ and the time step is $\dt=10^{-5}$. The unit of mass is set to one.
All quantities given below are expressed in these reduced units. This means that to obtain the numerical value of a length, it must be multiplied with $\dx$, and velocities have  a non-written factor of $\dx/\dt$. With this convention, the size of simulation domain is $L_x=L_y=800$ in 2D and $L_x=L_y=L_z=300$ in 3D. The numerical interface thickness is $\eta=6$ both in 2D and 3D simulations. The speed of sound is set to $\cs=0.12$ and constants of the van der Waals fluid, Eq.~(\ref{eq:EOS}), are set to $a=6.4$, $b=6000$, and $c=0.75\times10^{-6}$. For simplicity, we set viscosity of gas and liquid phases identical, $\mu=\mu_\textrm{Gas} = \mu_\textrm{Liq} = 1$. The initial radius of all bubble-nuclei is set to $R(t=0)=6$ both in 2D and 3D simulations. A 'newborn' gas bubble thus has a radius equal to the interface thickness. Gas densities within these bubbles are randomly assigned from the intervals $\rhoG(t=0) \in [2.6-2.9]$ in 2D and $\rhoG(t=0) \in [3.24-3.6]$ in 3D. Then, the amount of gas inside each bubble is increased via $M_\alpha(t) =\lambda t + M_\alpha(0)$, where $M_\alpha(0)$ is the initial mass of gas inside that bubble. Here, $\lambda$ is set to $0.32\times10^{-4}$ in 2D and $0.16\times10^{-4}$ in 3D. The density of liquid phase is kept constant through the simulation, $\rhoL = 5200$.

As a test of physical consistency, it is instructive to use the above parameters and estimate the speed of sound in the liquid phase. Using the van der Waals EOS in Eq.~(\ref{eq:EOS}), we obtain
\begin{equation}
c^2_\text{s,Liq}=\dfrac{\partial p}{\partial\rho}\Big|_{\rho=\rhoL}=\dfrac{ab}{(b-\rhoL)^2}-c=\dfrac{a}{b}\Big(\dfrac{b}{b-\rhoL}\Big)^2-c.
\label{eq:sound-speed-liq}
\end{equation}
Inserting the above values for $a,\;b,\;c$ and $\rhoL$, one obtains $c^2_\text{s,Liq}\approx 0.06$ and thus  $c_\text{s,Liq} \approx 0.245$, which is roughly twice the speed of sound in the gas phase. It is also noteworthy that the parameter $c$ has essentially no effect on the value obtained for $c_\text{s,Liq}$. This is a consequence of the fact that the last term in Eq.~(\ref{eq:EOS}) becomes important only at moderate densities relevant for condensation processes and has hardly an effect for phase behavior of a dense melt pool considered in the present study.

The liquid-gas surface free energy is set to $\sigmaLG=20.5$ in 2D and $10.25$ in 3D. For bubble-bubble interface energy, we use $\sigmaGG=2\sigmaLG(1+q)$, which is a special case of Eq.~(\ref{eq:non-wetting}). It must be emphasized here that the $q$-parameter influences the nucleation of coalescence in our model via two closely related mechanisms. On the one hand, a larger $q$ leads to a higher coalescence barrier via disjoining pressure, Eq.~(\ref{eq:Pidisj}). On the other hand, it increases $\sigmaGG$ and thus makes the approaching motion of two bubbles towards each other energetically unfavorable.

\subsection{Benchmark tests}

\begin{figure}
	\centering
	\includegraphics[height=6.5cm]{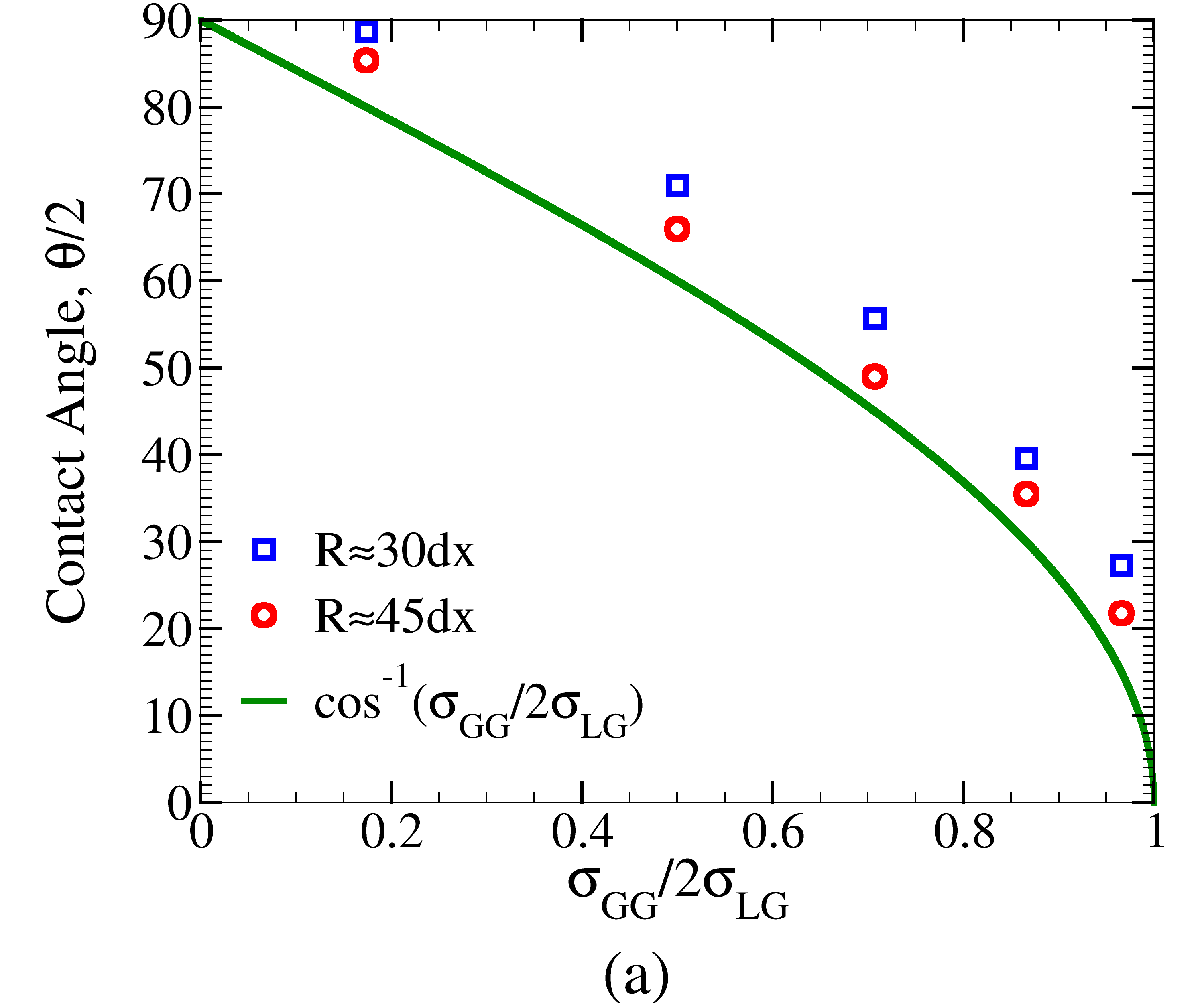}\hspace{1cm} \includegraphics[height=6.5cm]{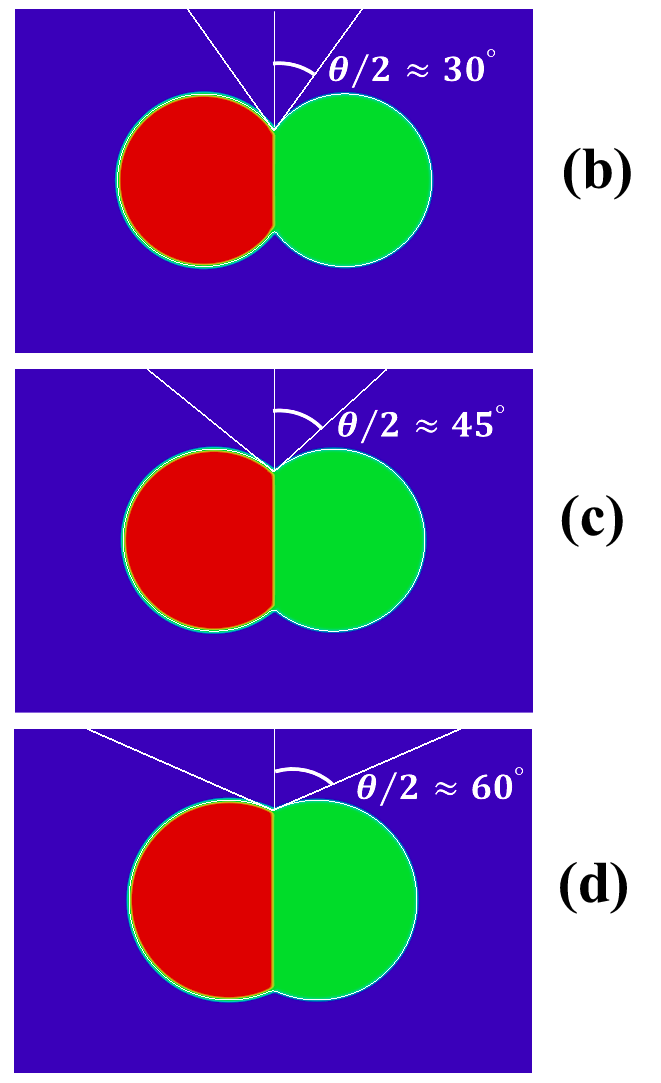}\\
	\caption{(a) Equilibrium contact angle, $\theta$, between two non-coalescing bubbles versus the ratio of surface energies, $\sigmaGG/\sigmaLG$. Symbols show simulations results for initial bubble radii of $R=30$ and $R=45$, and the solid line gives Eq.~(\ref{eq:contact-angle-sym}). For the larger bubble radius, simulations are closer to the theoretical predictions. This is expected because, as $R$ grows, the numerical evaluation of tangent lines at the triple junction becomes less biased by the interface curvature. (b-d) Equilibrium snapshots for small, intermediate and large values of $\theta$ for $R=45$.}
	\label{fig:theta}
\end{figure}

The first benchmark test deals with the static equilibrium between two non-coalescing bubbles. As discussed above, in such a situation, the contact angle satisfies Eq.~(\ref{eq:contact-angle-sym}). A validation of this relation is provided in Fig.~\ref{fig:theta}a, where simulation results for contact angle are plotted versus  $\sigmaGG/2\sigmaLG$. For each ratio of $\sigmaGG/2\sigmaLG$, two sets of simulations are performed using different bubble sizes. It is visible from this plot that larger bubbles reproduce the analytic result more closely. This is related to the fact that determination of angle between curved lines is more accurate if the radius of curvature is larger. In the opposite limit, one would have a systematic bias towards larger angles as curves go more quickly apart with distance from the crossing point. In agreement with this interpretation, it is seen from Fig.~\ref{fig:theta}a that angles obtained for smaller bubbles lie systematically above those for larger $R$.

\begin{figure}
	\centering
	\includegraphics[height=4.5cm]{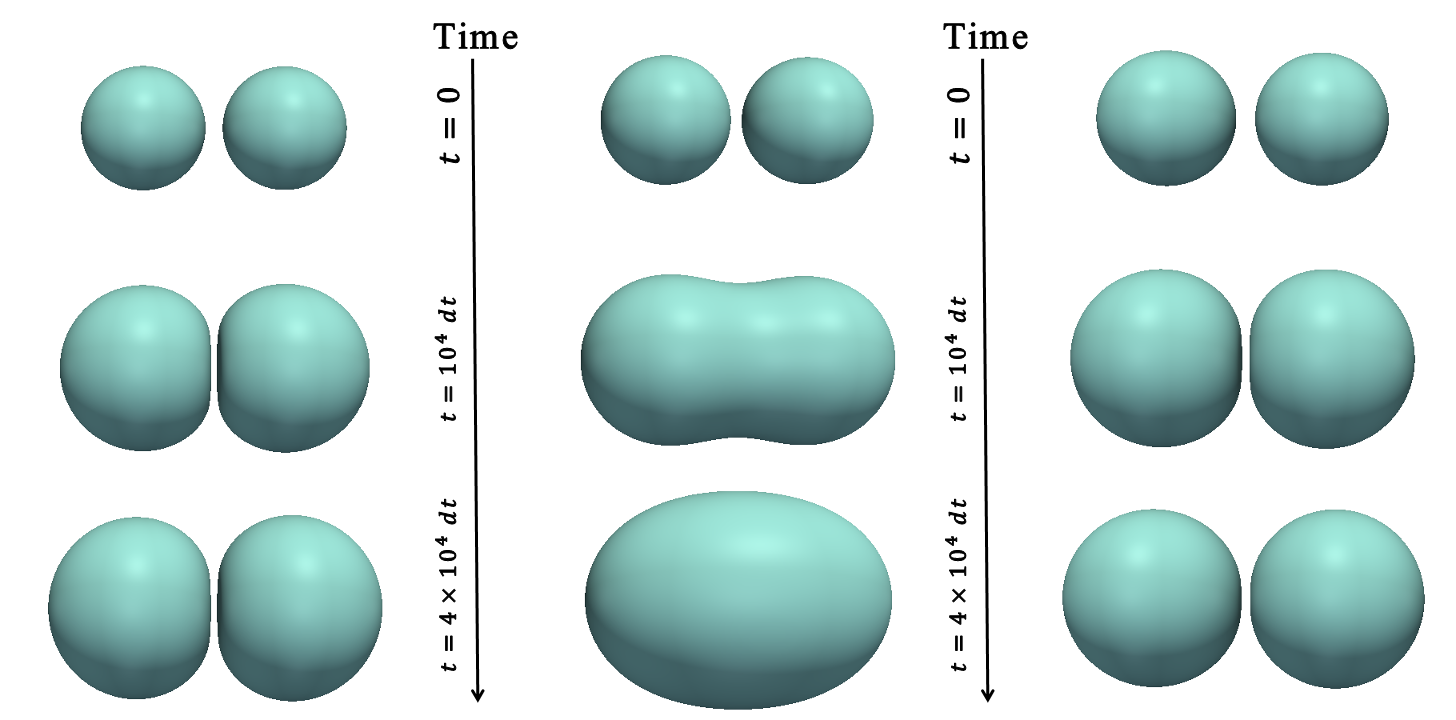}
	\includegraphics[height=4.5cm]{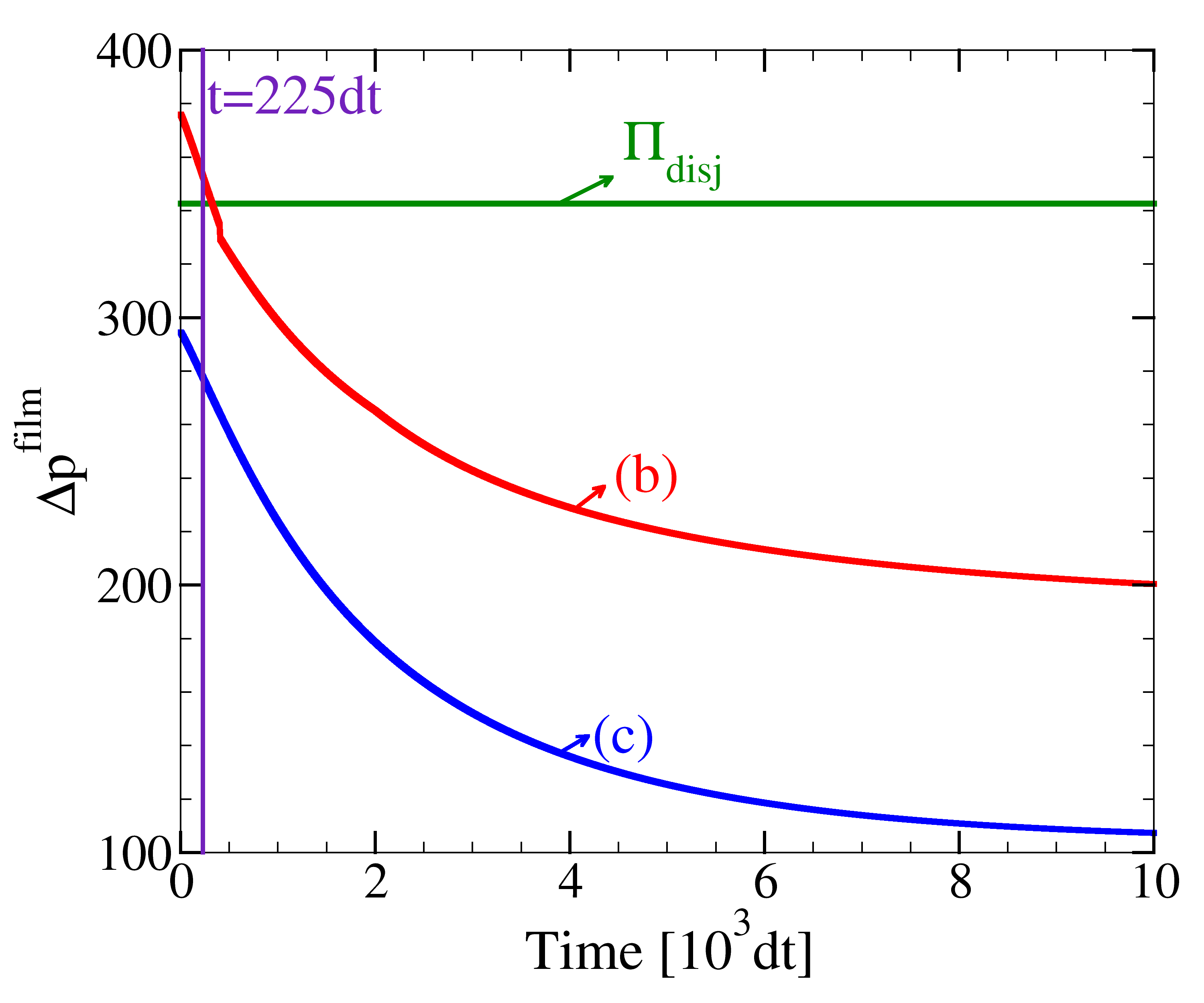}\\
	\hspace*{-0.07\textwidth}(a) \hspace{0.16\textwidth} (b) \hspace{0.165\textwidth} (c)\hspace{0.25\textwidth} (d)
	\caption{
		Test of the coalescence condition, inequality~(\ref{eq:coalescence-condition}). 
			In panel (a) bubbles remain separated throughout the entire process of growth and deformation because, in addition to a high gas-gas surface free energy, the phase fields representing the two bubbles are not allowed to merge into a single one. In (b) and (c), however, this possibility is introduced via the coalescence condition, inequality~(\ref{eq:coalescence-condition}). The only difference between (b) and (c) is the use of different initial bubble-densities. Otherwise, identical simulation parameters are employed. In (b) coalescence condition is satisfied at a time of $t\approx 225\dt$ and the bubbles merge into a single one (see also panel d). In (c), the initial densities within the bubbles are lower than in (b) so that driving forces always remain below the coalescence-threshold. (d) Variation of the $\dpfilm$ with time for the two cases (b) and (c). 
			The horizontal lines marks the threshold pressure, $\Pid$ which must be overcome for initiation of coalescence. In all the cases shown, the initial bubble radius is $R(t=0)=30\,=5\eta$.}
	\label{fig:2Bubbles}
\end{figure}

When two bubbles come into close contact, depending on forces which act upon them, they may undergo coalescence or remain separate. In the present model, the possibility for this bifurcation in dynamic behavior, which plays a key role in structure evolution in multi-bubble systems, is controlled via inequality (\ref{eq:coalescence-condition}). 
This issue is addressed in Fig.~\ref{fig:2Bubbles}, where two bubbles grow with time until they meet. In the first case shown (panel (a)), the non-wetting condition, Eq.~(\ref{eq:non-wetting}) is used with $q=1$. Apparently, the system minimizes its free energy by keeping the bubbles apart. Starting with the same configuration as in panel (a), simulations are repeated using the coalescence condition, inequality~(\ref{eq:coalescence-condition}). Two very similar cases that differ only in the initial gas density are considered. In one of these cases, the two bubbles merge and form a bigger one (Fig.~\ref{fig:2Bubbles}b). In the other case, which started with a lower gas density, bubbles remain separated during the entire simulation time (Fig.~\ref{fig:2Bubbles}c). This different behavior can be rationalized by a survey of the force, $\Delta p^\textrm{film}$, which drives the coalescence process. As seen in Fig.~\ref{fig:2Bubbles}d, in the coalescing case, there is a time where $\Delta p^\textrm{film}$ exceeds the disjoining pressure. In the setup corresponding to Fig.~\ref{fig:2Bubbles}c, however, the initial gas density inside bubbles is low and the driving force for coalescence remains below the threshold during the whole simulation. 

\begin{figure}
\centering
\includegraphics[width=0.45\linewidth]{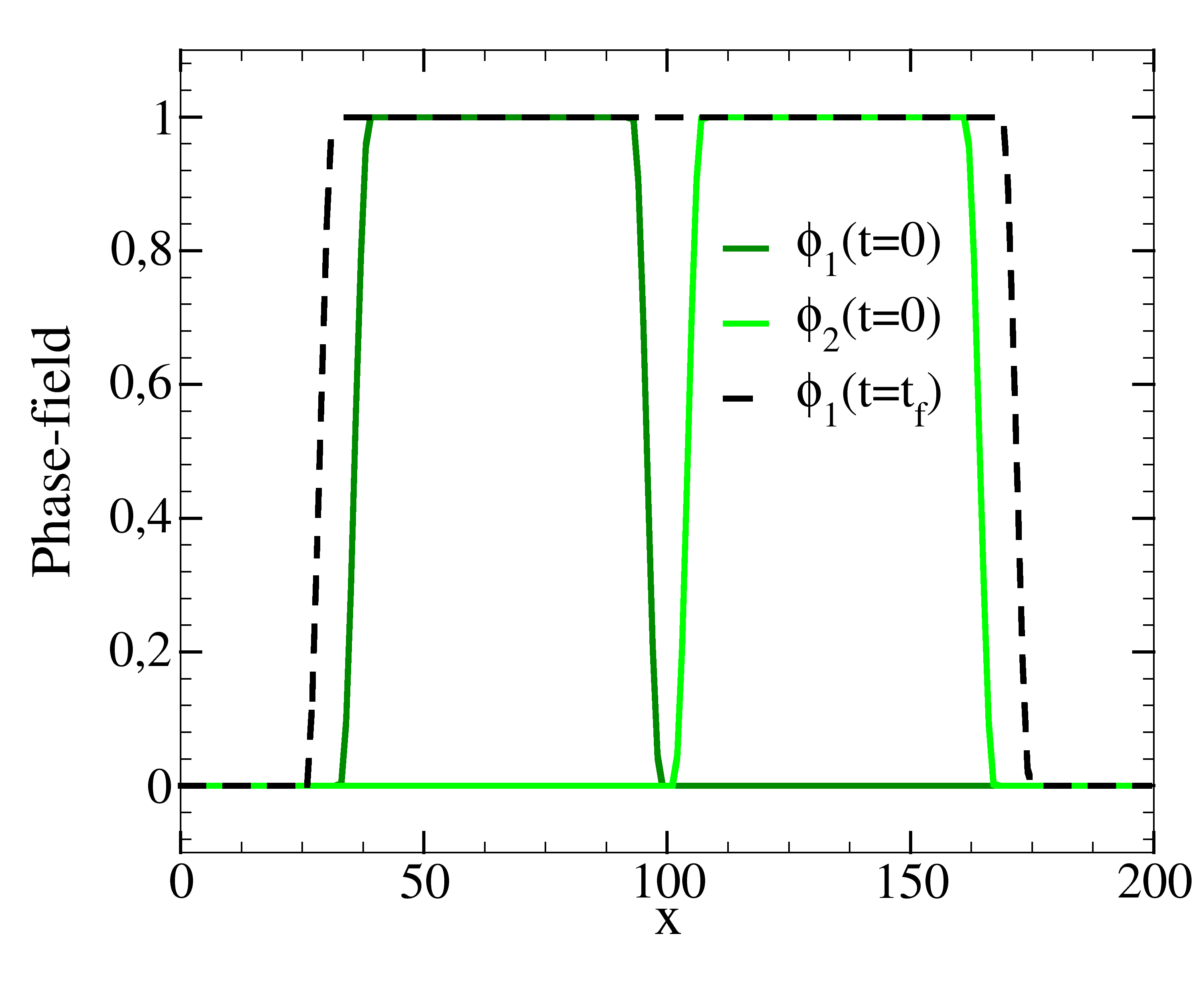} \includegraphics[width=0.45\linewidth]{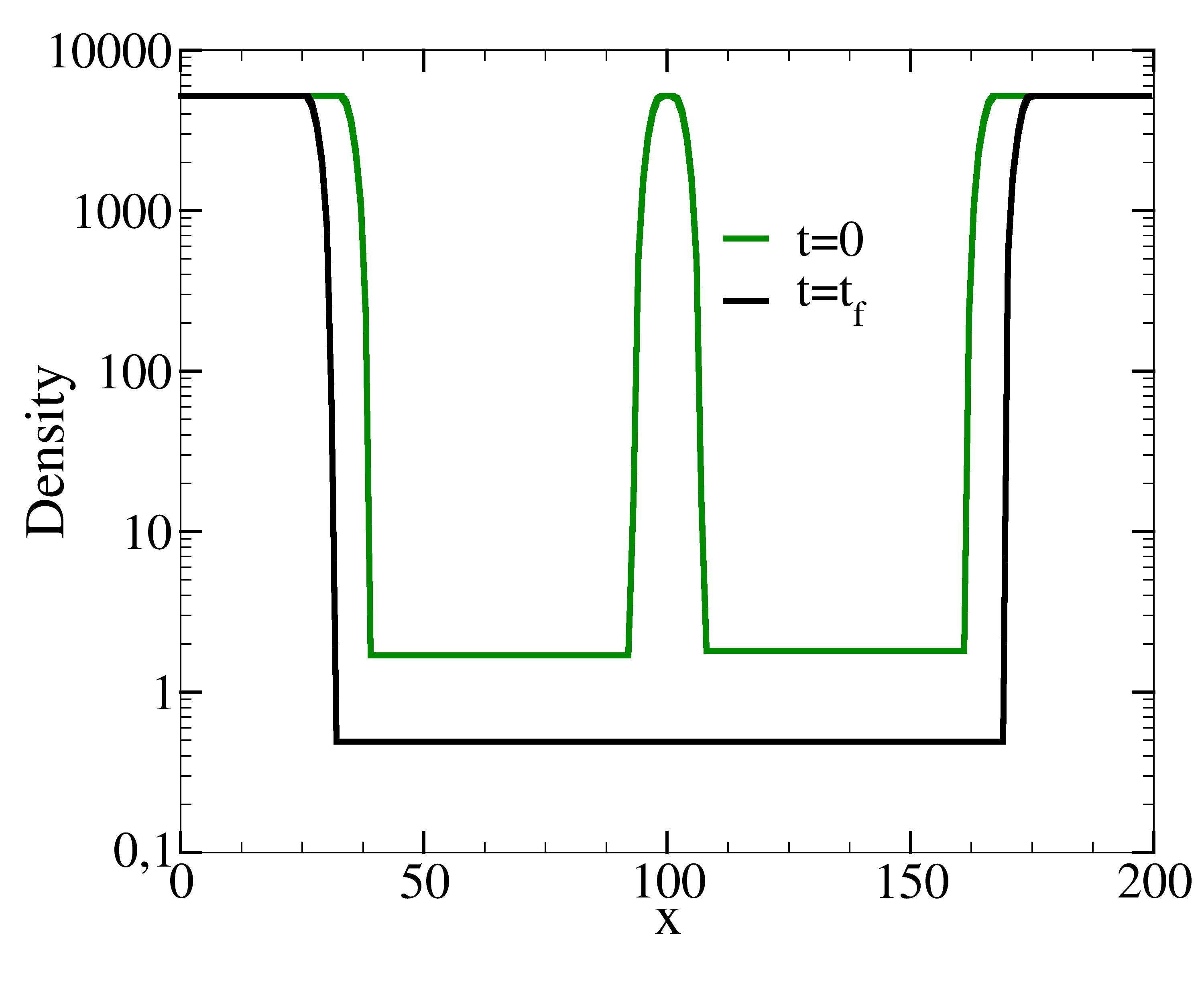}\\
\centering (a) \hspace{0.42\textwidth} (b)
\caption{ (a) The phase-field and (b) density profiles corresponding to the system of bubbles in Fig.~\ref{fig:2Bubbles}b. These profiles are plotted along the horizontal line passing through the center of the bubbles in Fig.~\ref{fig:2Bubbles}b. The phase-field profiles of the bubbles are shown in dark and light green for the initial time step, $t=0$, and in black for final time step, $t=t_f$, in (a). Similarly, the green and black lines in (b) correspond to the density profile at time $t=0$ and $t=t_f$.}
\label{fig:2Bubbles-profile}
\end{figure}

In the instant of coalescence in Fig.~\ref{fig:2Bubbles}b, one of the two phase-fields is deleted and the entire gas domain is assigned to the other one. This is shown in Fig.~\ref{fig:2Bubbles-profile}a, where the green lines represent the phase-field profiles of the bubbles at time $t=0$ and the black line corresponds to the final state of the single bubble, which is the product of coalescence. All the profiles are plotted along the horizontal line passing through the center of coalescing bubbles. Density profiles before and after coalescence are also shown in Fig.~\ref{fig:2Bubbles-profile}b.

The results discussed above clearly show that the present model has the capability to adequately account for static equilibrium and dynamic behavior of bubbles. Most importantly, it provides a physically motivated model to study structure formation in foams by completely preventing bubble-coalescence. At the same time, the model also contains a barrier-controlled criterion for coalescence, which can be used to tune the rate of coalescence and the resulting microstructure. 
Next, we show the results of many-bubble simulations both in two and three dimensions which underline the maturity of our model in dealing with complex structures. 

\subsection{Many-bubble simulations}

\begin{figure}
\centering
\includegraphics[width=4.8cm]{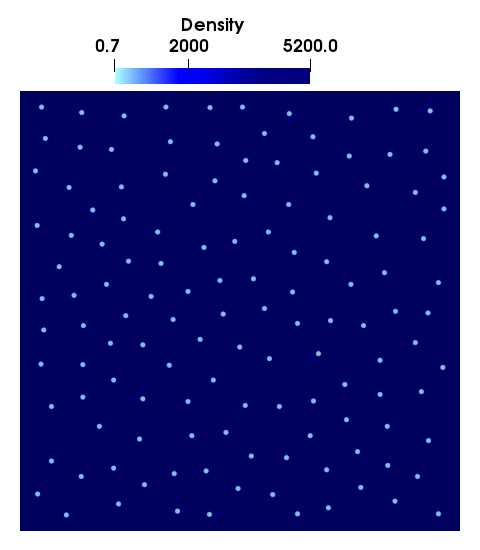}
\includegraphics[width=4.8cm]{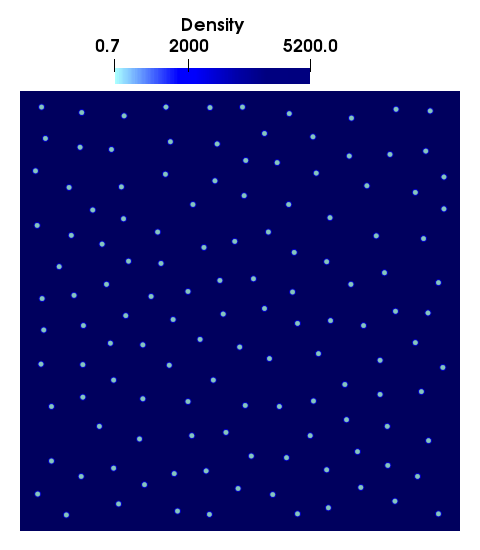}
\includegraphics[width=4.8cm]{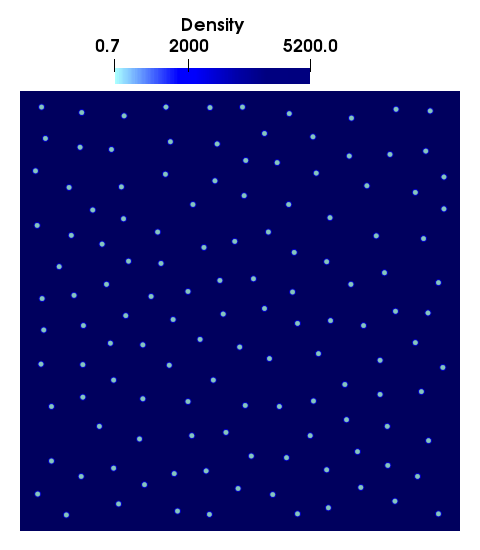}\\
\includegraphics[width=4.8cm]{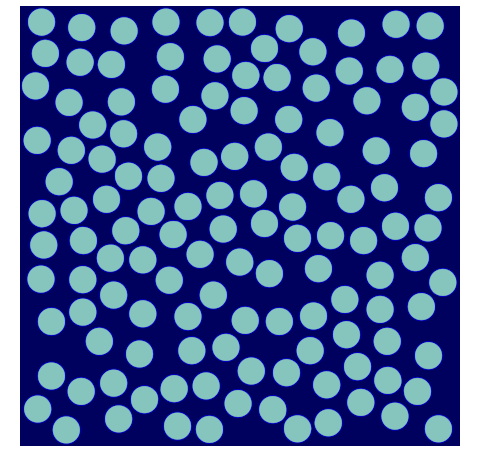}
\includegraphics[width=4.8cm]{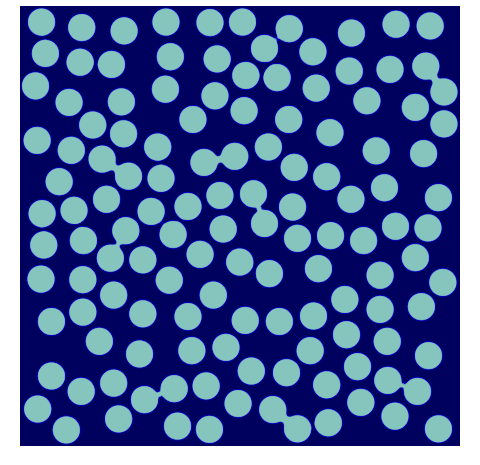}
\includegraphics[width=4.8cm]{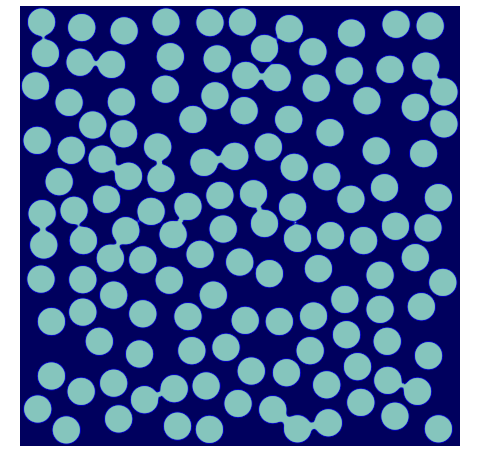}\\
\includegraphics[width=4.8cm]{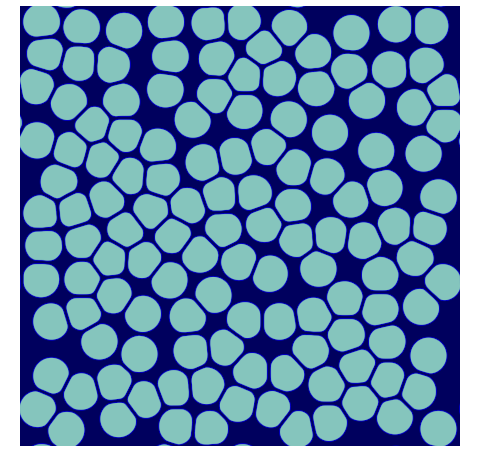}
\includegraphics[width=4.8cm]{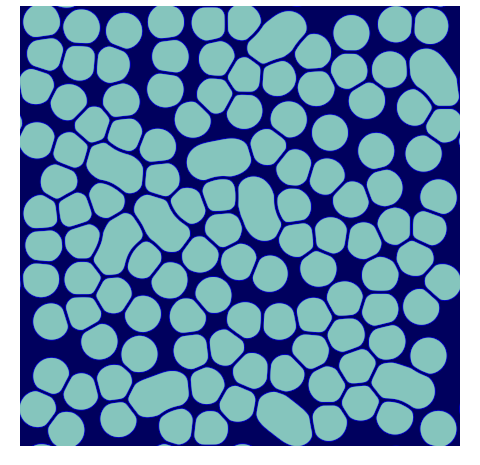}
\includegraphics[width=4.8cm]{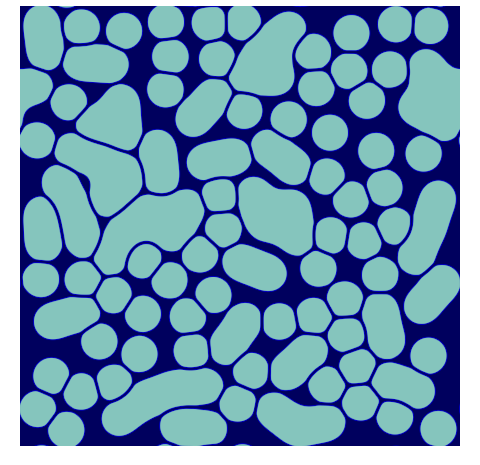}\\
\includegraphics[width=4.8cm]{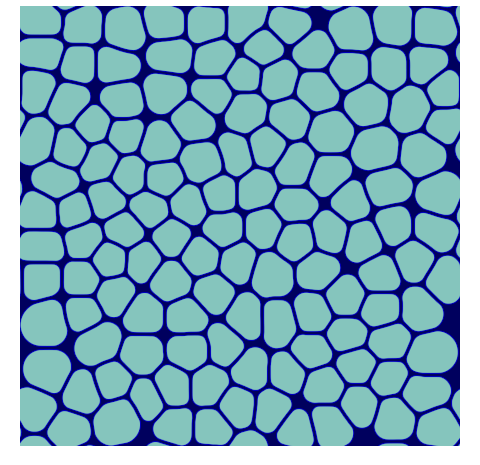}
\includegraphics[width=4.8cm]{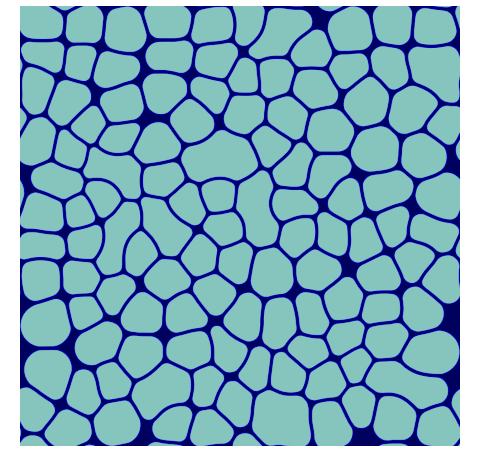}
\includegraphics[width=4.8cm]{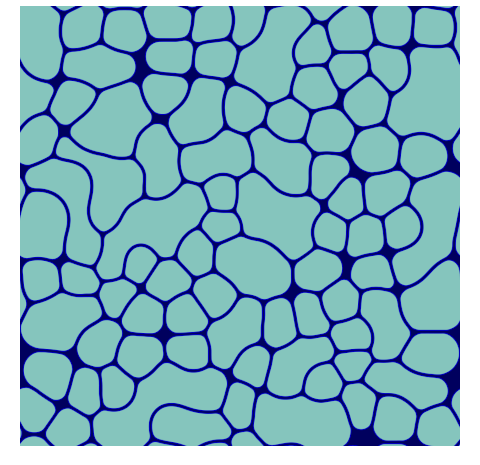}\\
(a) \hspace*{0.28\textwidth} (b) \hspace*{0.28\textwidth} (c)\\
\caption{Time evolution (from top to bottom) of many bubbles and the resulting foam structure for different surface free energies and disjoining pressures by tuning the $q$ parameter in Eqs.~(\ref{eq:non-wetting}) and~(\ref{eq:Pidisj}), (a) $q=0.25$, (b) $q=0.24$, (c) $q=0.23$. The color code represents the density field, dark blue for the liquid phase and light blue for the gas phase. From top to bottom, each row corresponds to $t=0$, $t=0.9\times 10^4\dt$, $t=1.5\times 10^4\dt$, and $t=10^5\dt$, respectively. The system size is $800\times 800$ in the units of grid spacing, the interface width is set to $\eta=6\hspace{0.5mm}\dx$, and $\rhoL/\rhoG \approx 10,000$.}
\label{fig:Foam2D}
\end{figure}

\begin{figure}
\centering
\includegraphics[width=0.5\linewidth]{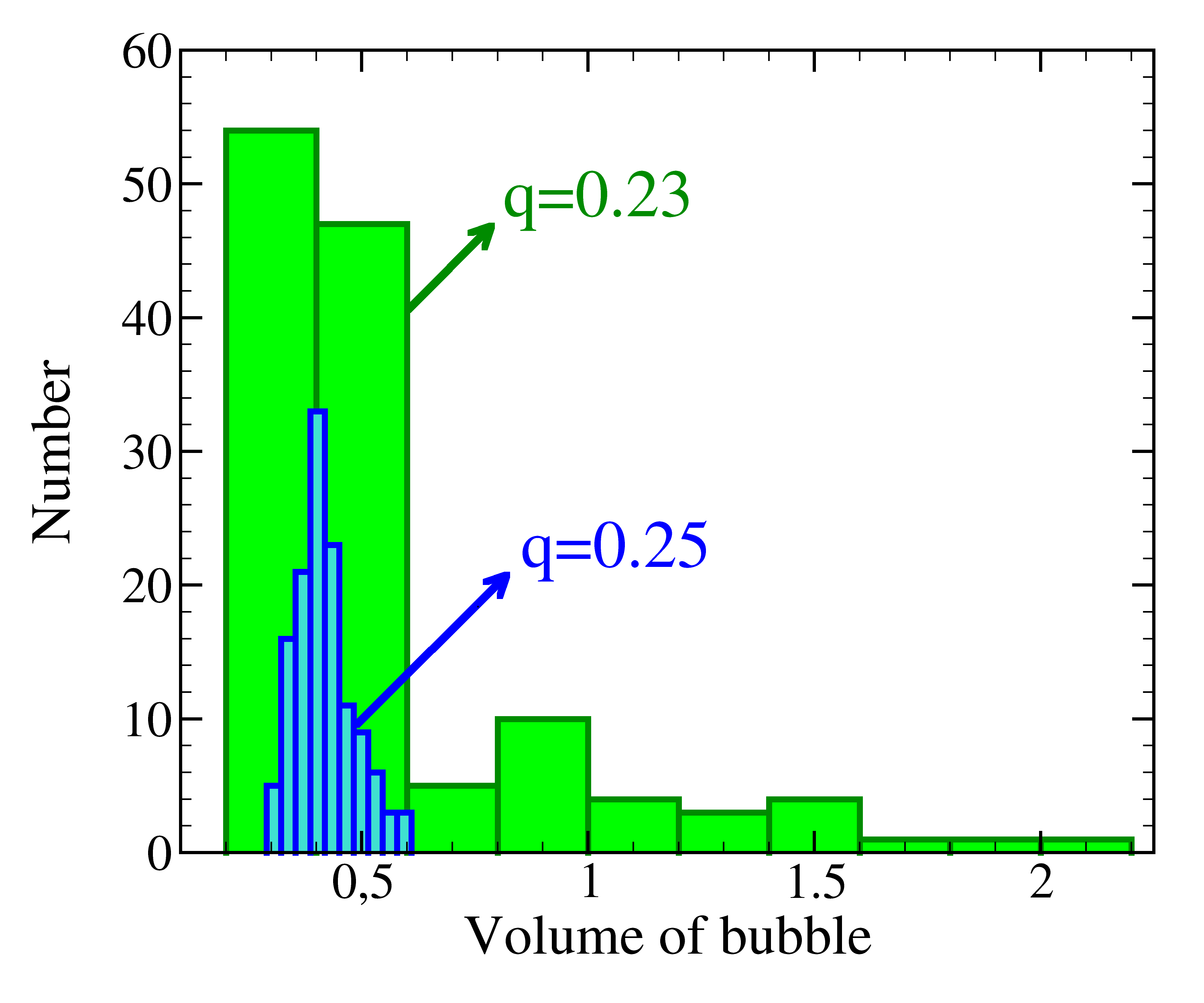}
\caption{Histogram of bubbles volume distribution at time step $t=10^5 \dt$ for foam structure using $q=0.23$ (green) and $0.25$ (blue). The size of bubbles are more homogeneous for the case of $q=0.25$.}
\label{fig:Histogram}
\end{figure}

We employ the model to simulate formation of foam microstructures for different disjoining pressures. This is achieved by changing the value of $q$ in Eqs.~(\ref{eq:non-wetting}) and~(\ref{eq:Pidisj}). Since disjoining pressure determines the coalescence barrier, Eq.~(\ref{eq:coalescence-condition}), the number of bubbles which merge will vary with $q$. This, in turn, will affect the number of pores and their size distribution, leading to variable foam structures. Here, we performed a set of simulations using three different values of $q$, $q_1=0.25$, $q_2=0.24$, and $q_3=0.23$. Except for the value of $q$, all the other parameters and conditions were identical in the three simulations reported below. In all the cases shown, bubble nuclei grow due to the increase of mass until they come into contact, deform and rearrange, Fig.~\ref{fig:Foam2D}a-c. The process continues until the system is filled with bubbles separated by the liquid films (see the last row in Fig.~\ref{fig:Foam2D}a-c). For the case of $q=0.25$, coalescence is hindered most effectively and the final structure contains a homogeneous distribution of pores, Fig.~\ref{fig:Foam2D}a. For the smaller value of $q=0.24$ (Fig.~\ref{fig:Foam2D}b), coalescence is less suppressed and, consequently, the number of pores is decreased. In this case, distribution of bubble size is broadened since larger pores are produced due to coalescence. This results in a less homogeneous foam structure, as compared to the case of larger disjoining pressure (cf. panels a and b in Fig.~\ref{fig:Foam2D}). For even lower $\Pi_\text{disj}$ ($q=0.23$), still more bubbles merge into each other and this creates a foam including pores with a very large range of size, Fig.~\ref{fig:Foam2D}c. These results suggest that homogeneity of pore structure is deteriorated by lowering the coalescence barrier. Moreover, as highlighted in Fig.~\ref{fig:Histogram}, this loss of homogeneity is accompanied by a broader bubble size distribution.

\begin{figure}
\centering
\includegraphics[height=6cm]{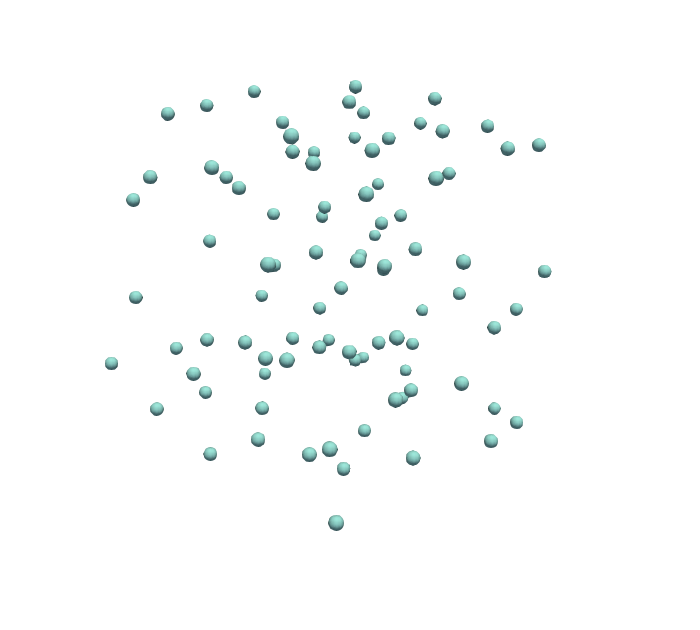}
\includegraphics[height=6cm]{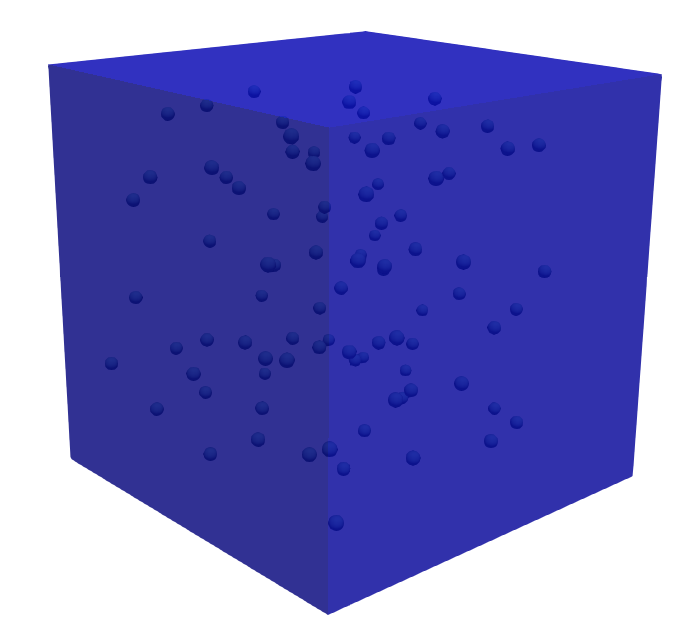}\\
\includegraphics[height=6cm]{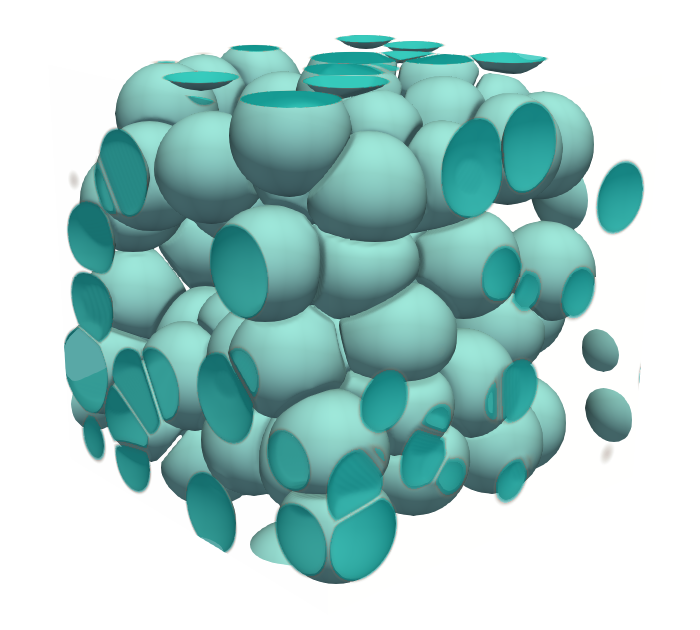}
\includegraphics[height=6cm]{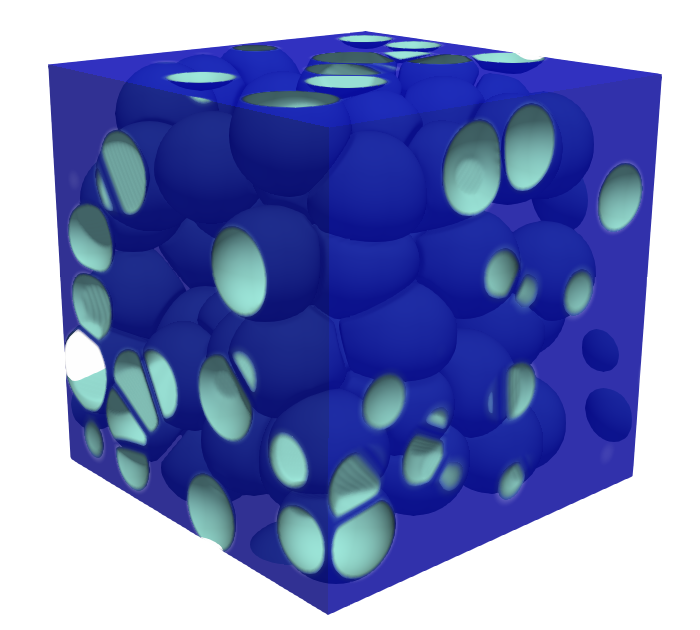}\\
\includegraphics[height=6cm]{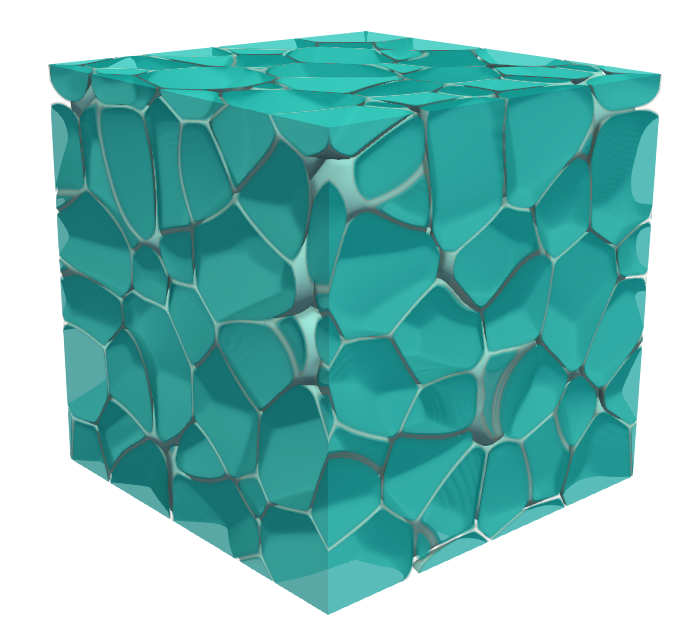}
\includegraphics[height=6cm]{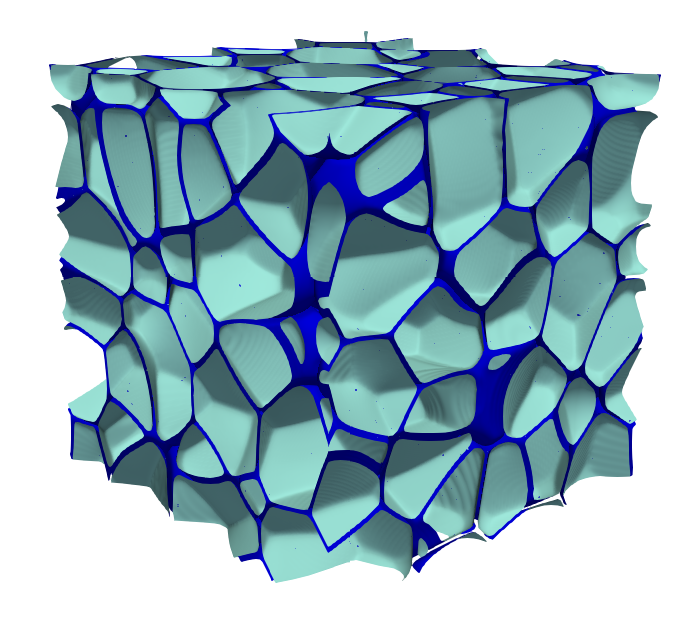}\\
(a) \hspace{0.45\linewidth}(b)
\caption{Result of 3D simulation for microstructure evolution of a foam representing bubbles in (a) and the liquid film around them in (b). Each raw, beginning from the top, corresponds to $t=0$, $1.5\times10^4$, and $5\times10^4dt$. The system size is $300\times300\times300$ in the units of numerical resolution. The interface width is set to $\eta=6\hspace{0.5mm}dx$. The initial density of each bubble is assigned from the range of  $\rho_\alpha(t=0) = 3.24-3.6$ and the density of the liquid is constant throughout the simulation, $\rho_\textrm{l}$. Thus, the density ratio is around $\rho_\textrm{l}/\rho_\textrm{g} \approx 10,000$.}
\label{fig:Foam3D}
\end{figure}

As the last example, we provide results on a 3D foam-microstructure simulated with the present model. To obtain a homogeneous structure and based on the knowledge gained from the above 2D simulations, we choose a very high free energy barrier for coalescence by setting $q=1$ in Eqs.~(\ref{eq:non-wetting}) and~(\ref{eq:Pidisj}). The result of this simulation is shown in Fig.~\ref{fig:Foam3D}. The left column corresponds to bubbles/pores which evolve as a result of mass addition until they fill the simulation box. During the growth process, as the distance between two adjacent bubbles falls below $\eta$, the energy barrier prevents their coalescence. Thus, bubbles deform by growing further until the structure reaches a semi-stable configuration (last row in Fig.~\ref{fig:Foam3D}).

\section{Conclusion}
In this paper, we represent a 3D multi-phase-field model for simulation of microstructure evolution in metallic foams. First, a formulation of the model is presented which allows to completely avoid coalescence between bubbles. The model is then extended by a rule to allow for coalescence if (i) the liquid film between the bubble becomes thinner than a threshold and (ii) the force, which pushes the bubbles towards each other, surpasses the disjoining pressure. This approach is validated through benchmark tests and is shown to reproduce the analytically expected results. The thus established model is used to study structure evolution is two and three-dimensional cases. The present model provides a powerful and versatile numerical approach to study foam formation under various physical and process conditions. A direction of particular practical interest would be to explore various gas production mechanisms in order to obtain high surface-to-volume ratio, low density, and high mechanical properties. Moreover, the use of ideal-gas equation of state for bubbles is an idealization that can be easily replaced by a more realistic one in future studies.

\section*{Acknowledgments}
This work was performed with support from the IMPRS-SurMat program.

\appendix
\begin{appendices}

\numberwithin{equation}{section}

\section{Derivation of pressure tensor}
\label{appSec:derivation of pressure tensor}

In order to obtain a closed expression for the pressure tensor, instead of $\cal F$, we start from a slightly different free energy functional,
\begin{equation}
\begin{aligned}
\tilde{\mathcal{F}} &= \int_{\Omega}^{} \Big(\mathcal{L} \big( \{ \phi \}, \{ \nabla \phi \} \big) +\lambda(t) \big( \sum_{\alpha=1}^{N} \phi_\alpha -1 \big) \Big) dV,
\end{aligned}
\label{app::eq:Ftilde}
\end{equation}
where $\lambda(t)$ is a Lagrange multiplier. As seen from a comparison with Eq.~(\ref{eq:F}), the integral over the first term is the usual free energy functional, $\cal F$. The second term in Eq.~(\ref{app::eq:Ftilde}) ensures that the sum of all phase fields is conserved at any point in space.

To simplify notation in the following derivation, we introduce $\tilde{\mathcal{L}} \equiv \mathcal{L} \big( \{ \phi \}, \{ \nabla \phi \} \big) +\lambda(t) \big( \sum_{\alpha=1}^{N} \phi_\alpha -1 \big)$. With this definition, Eq.~(\ref{app::eq:Ftilde}) takes the form
\begin{eqnarray}
\tilde{\mathcal{F}} = \int_{\Omega}^{} \tilde{\mathcal{L}} \big( \{ \phi \}, \{ \nabla \phi \} \big) dV.
\label{app::eq:Ftilde-Ltilde}
\end{eqnarray}
At equilibrium,
\begin{equation}
\frac{\partial \tilde{\mathcal{L}}}{\partial \phi_{\alpha}} - \frac{d}{d x_{i}} \frac{\partial \tilde{\mathcal{L}}}{\partial \partial_{i} \phi_{\alpha}} = 0,
\label{eq:EL-Ltilde}
\end{equation}
for $\alpha=1,...,N$. Similar to \cite{Vakili2020}, the total derivative for $\tilde{\mathcal{L}}$ yields
\begin{eqnarray}\nonumber
\frac{d \tilde{\mathcal{L}}}{d x_{j}} &=& \frac{\partial \tilde{\mathcal{L}}}{\partial x_{j}} + \sum_{\alpha=1}^{N} \Big( \frac{\partial \tilde{\mathcal{L}}}{\partial \partial_{i} \phi_{\alpha}} \partial_{j} \partial_{i} \phi_{\alpha} + \frac{\partial \tilde{\mathcal{L}}}{\partial \phi_{\alpha}} \partial_{j} \phi_{\alpha}\Big) \\
&=& \frac{\partial \tilde{\mathcal{L}}}{\partial x_{j}} + \sum_{\alpha=1}^{N} \frac{d }{dx_{j}} \Big(\frac{\partial \tilde{\mathcal{L}}}{\partial \partial_{i} \phi_{\alpha}} \partial_{j} \phi_{\alpha}\Big),
\label{app_total_derivative}
\end{eqnarray}
where $\partial \tilde{\mathcal{L}}/\partial \phi_{\alpha}$ is replaced by $(d/d x_{i}) (\partial \tilde{\mathcal{L}}/\partial \partial_{i} \phi_{\alpha})$ and then the product rule is used to derive the second line. Equation~(\ref{app_total_derivative}) is reordered to
\begin{equation}
\frac{\partial \tilde{\mathcal{L}}}{\partial x_{j}} = \frac{d}{d x_{i}} \bigg(  \tilde{\mathcal{L}} \delta_{ij} - \sum_{\alpha =1}^{N-1} \frac{\partial \tilde{\mathcal{L}}}{\partial \partial_{i} \phi_{\alpha}} \partial_{j} \phi_{\alpha} \bigg) \equiv \nabla \cdot \mathbf{P}.
\label{app:total_derivative2}
\end{equation}
Similar to pressure, the tensor $\mathbf{P}$ introduced above has the dimension of energy density. Moreover, it follows from Eq.~(\ref{app:total_derivative2}) that this tensor is divergence free if the Lagrange function $\cal{L}$ obeys translational invariance, i.e., if it does not explicitly depend on position. The close connection between momentum conservation and translational invariance then implies the existence of a divergence free pressure tensor, which we identify as $\mathbf{P}$~\cite{Goldstein1980}. To proceed further, we express $\tilde{\mathcal{L}}$ in terms of $\cal{L}$ and Lagrange multiplier to obtain ($\mathbf{I}$ is the identity tensor),
\begin{equation}
\nabla \cdot \mathbf{P} = \nabla \cdot \Big(\mathcal{L}\;\mathbf{I}  - \sum_{\alpha=1}^{N}  \frac{\partial \mathcal{L}}{\partial \nabla \phi_{\alpha}} \nabla \phi_{\alpha} \Big)  + \lambda(t) \sum_{\alpha=1}^{N} \nabla\phi_{\alpha}.
\label{app:interfacial_force}
\end{equation}
The Lagrange multiplier, $\lambda(t)$, can be expressed via derivatives of $\cal{L}$ by using the Euler-Lagrange equations~(\ref{eq:EL-Ltilde}). The result is
\begin{equation}
\lambda(t) = -\frac{1}{N}\sum_{\beta=1}^{N} \Big( \frac{\partial \mathcal{L}}{\partial \phi_{\beta}} - \nabla \cdot\frac{\partial \mathcal{L}}{\partial \nabla \phi_{\beta}} \Big).
\label{app:lambda}
\end{equation}
Inserting this relation in Eq.~(\ref{app:interfacial_force}) yields
\begin{eqnarray}\nonumber
\nabla \cdot \mathbf{P} &=& \frac{N}{N} \sum_{\alpha=1}^{N}  \Big( \frac{\partial \mathcal{L}}{\partial \phi_{\alpha}} \nabla \phi_\alpha - \big(\nabla \cdot \frac{\partial \mathcal{L}}{\partial \nabla \phi_{\alpha}}\big) \nabla \phi_{\alpha} \Big)  - \frac{1}{N} \sum_{\alpha=1}^{N}\sum_{\beta=1}^{N} \Big( \frac{\partial \mathcal{L}}{\partial \phi_{\beta}} - \nabla \cdot \frac{\partial \mathcal{L}}{\partial \nabla \phi_{\beta}} \Big)  \nabla\phi_{\alpha}\\
&=& \frac{1}{N} \sum_{\alpha=1}^{N}\sum_{\beta=1}^{N} \bigg\{ \Big( \frac{\partial \mathcal{L}}{\partial \phi_{\alpha}} - \nabla \cdot \frac{\partial \mathcal{L}}{\partial \nabla \phi_{\alpha}} \Big)  -  \Big( \frac{\partial \mathcal{L}}{\partial \phi_{\beta}} - \nabla \cdot \frac{\partial \mathcal{L}}{\partial \nabla \phi_{\beta}} \Big) \bigg\} \nabla\phi_{\alpha}\nonumber\\
&=& \frac{1}{N}\sum_{\alpha=1}^{N}\sum_{\beta=1}^{N} \bigg\{ \frac{\delta \mathcal{F}}{\delta \phi_{\alpha}} - \frac{\delta \mathcal{F}}{\delta \phi_{\beta}} \bigg\} \nabla\phi_{\alpha}.
\label{eq:divP2}
\end{eqnarray}

In order to evaluate forces arising from divergence of the pressure tensor in terms of the model parameters, the free energy density, $\cal L$, must be specified. A standard choice is~\cite{Vakili2020}
\begin{equation}
\mathcal{L} = \sum_{\alpha = 1}^{N-1} \sum_{\beta = \alpha+1}^{N} \Big( -\frac{W^2_{\alpha\beta}}{2} \nabla \phi_{\alpha} \cdot \nabla \phi_{\beta} +  \frac{\gamma_{\alpha \beta}}{2} |\phi_{\alpha} \phi_{\beta}| - \big[h(\phi_{\alpha}) p_{\alpha}(\rho_{\alpha}) + h(\phi_{\beta}) p_{\beta}(\rho_{\beta})\big] \Big),
\label{eq:L-app}
\end{equation}
where $W^2_{\alpha \beta}$ is the analog of square-gradient coefficient for the case of multiple phases, $\gamma_{\alpha\beta}$ tunes the strength of potential energy and $p_\alpha$ and $p_\beta$ are bulk pressures within the phases $\alpha$ and $\beta$, respectively. The sum in square brackets is an average pressure with $h$ playing the role of an interpolation function. 
It is possible to express $W^2_{ij}$ and $\gamma_{ij}$ in terms of interface energy and thickness, $\sigma_{\alpha \beta}$ and $\eta$, respectively. For this purpose, we consider force balance at a planar interface and obtain (see Appendix~\ref{appSec:planar}),
\begin{eqnarray} 
\label{eq:gamma}
\gamma_{\alpha \beta} &= \dfrac{8\sigma_{\alpha \beta}}{\eta},\\
W^2_{\alpha \beta}&= \dfrac{8\sigma_{\alpha \beta} \eta }{\pi^2}.
\label{eq:W}
\end{eqnarray}
Inserting Eqs.~(\ref{eq:gamma}) and ~(\ref{eq:W}) in Eq.~(\ref{eq:L-app}), one arrives at
\begin{equation}
\mathcal{L} = \sum_{\alpha = 1}^{N-1} \sum_{\beta = \alpha+1}^{N} \Big( -\frac{4 \sigma_{\alpha\beta}\eta}{\pi^2} \nabla \phi_{\alpha} \cdot \nabla \phi_{\beta} +  \frac{4\sigma_{\alpha \beta}}{\eta} \phi_{\alpha} \phi_{\beta} - h(\phi_{\alpha}) p_{\alpha}(\rho_{\alpha}) - h(\phi_{\beta}) p_{\beta}(\rho_{\beta}) \Big).
\label{eq:L2}
\end{equation}
Guidance for a reasonable choice of the interpolation function $h$ can be obtained by considering a spherical bubble in equilibrium with the surrounding liquid. One can then verify that the following choice of the function $h$ satisfies the force balance condition for a spherical bubble (see Appendix~\ref{appSec:sphere}),
\begin{equation}
h(\phi_{\alpha}) = \frac{1}{\pi}\Big(2(2\phi_\alpha-1)\sqrt{\phi_\alpha(1-\phi_\alpha)} + \arcsin(2\phi_\alpha-1) + \frac{\pi}{2} \Big).
\label{Interpolation_function}
\end{equation}

Using Eq.~(\ref{eq:L2}), divergence of the pressure tensor can be expressed in terms of model parameters,
\begin{eqnarray}\nonumber
\nabla \cdot \mathbf{P} &=&  \sum_{\alpha=1}^{N}\sum_{\beta=1}^{N}  \frac{1}{N} \bigg\{ \Big(\sum_{\xi=1,\xi \neq \alpha}^{N}-\frac{4\sigma_{\alpha\xi}\eta}{\pi^2}\big(\frac{\pi^2}{\eta^2}\phi_\xi+\nabla^2\phi_\xi\big)+\sum_{\xi=1,\xi \neq \beta}^{N}\frac{4\sigma_{\beta\xi}\eta}{\pi^2}\big(\frac{\pi^2}{\eta^2} \phi_\xi+\nabla^2\phi_\xi\big)\Big)  +\\
&& \nonumber(N-1)\Big(p_\alpha\frac{\partial h}{\partial\phi_\alpha}-p_\beta\frac{\partial h}{\partial\phi_\beta} \Big)\bigg\} \nabla\phi_{\alpha}\\
&=&\frac{4\eta}{\pi^2 N} \sum_{\alpha=1}^{N}\sum_{\beta=1}^{N} \bigg\{ \sum_{\xi=1}^{N}(\sigma_{\beta\xi}-\sigma_{\alpha\xi})I_\xi +\frac{(N-1)\pi^2}{4\eta}\Big(p_\alpha\frac{\partial h}{\partial\phi_\alpha}-p_\beta\frac{\partial h}{\partial\phi_\beta} \Big)\bigg\} \nabla\phi_{\alpha}.
\label{eq:divP3}
\end{eqnarray}
In Eq.~(\ref{eq:divP3}), $I_\xi =\nabla^2\phi_\xi+(\pi\phi_\xi/\eta)^2$ reflects curvature effects \cite{Vakili2017}. The second bracket on the rhs of Eq.~(\ref{eq:divP3}) is the contribution of hydrostatic pressure to the driving force. 

Equation~(\ref{eq:divP3}) becomes relatively simple for the special case of a two-phase system. With $\phi_\alpha=\phi$ and $\phi_\beta=1-\phi$ one then obtains,
\begin{equation}
\nabla \cdot \mathbf{P} = \Big( (p_\alpha-p_\beta)\dfrac{\partial h}{\partial \phi} - \gamma_{\alpha\beta}(\frac{1}{2}-\phi) + W_{\alpha\beta}^2 \nabla^2\phi \Big) \nabla \phi. \;\;\;\;\textrm{(two-phase system)}
\label{eq:divP4}
\end{equation}
Note that the interpolation function $h$ appears only in a product with pressure difference between the two phases. As a consequence, $h$ plays no role in force balance if hydrostatic pressure does not vary across the interface.
An example is the equilibrium condition for two phases separated by a planar interface, which we discuss below.

\subsection{Planar interface}
\label{appSec:planar}
For a planar interface, it directly follows from the force balance condition at equilibrium ($\nabla\cdot\mathbf{P}=\mathbf{0}$) that the component of pressure tensor along the direction normal to the interface is spatially constant. This implies $p_\alpha=p_\beta$~\cite{Vakili2020}.
Inserting this information into the rhs of Eq.~(\ref{eq:divP4}), the term containing the interpolation function drops and one obtains,
\begin{equation}
-  \gamma_{\alpha\beta}  \Big(\dfrac{1}{2}-\phi\Big) +   W_{\alpha\beta}^2 \dfrac{\partial^2\phi}{\partial x^2}= 0,
\label{eq:static-equil-planar}
\end{equation}
where we assumed that the interface is normal to the $x$-direction.  
For the double obstacle potential used in the present study, the interface profile is given by~\cite{Steinbach2009}, 
\begin{equation}
\phi(x) =
\begin{cases}
1 & \hspace{10mm}x \le - \dfrac{\eta}{2}\\
\dfrac{1}{2} - \dfrac{1}{2}\sin\Big(\dfrac{\pi}{\eta}x\Big) & -\dfrac{\eta}{2} \le x \leq \dfrac{\eta}{2} \\
0 & \hspace{10mm}x \geq \dfrac{\eta}{2},
\end{cases}
\label{eq:sin-profile}
\end{equation}
where $\eta$ is the interface width. Substitution of Eq.~(\ref{eq:sin-profile}) into Eq.~(\ref{eq:static-equil-planar}) gives
\begin{equation}
\gamma_{\alpha\beta}= \dfrac{\pi^2W^2_{\alpha\beta}}{\eta^2}.
\label{eq:gamma-ab}
\end{equation}
On the other hand, the specific interface free energy is given by integral over the free energy density, $\mathcal{L}$, Eq.~(\ref{eq:L-app}), excluding the pressure term which contains the bulk free energy. It reads
\begin{equation}
\sigma_{\alpha\beta}= \int_{-\infty}^{\infty} \bigg(\dfrac{W^2_{\alpha\beta}}{2}\Big(\dfrac{\partial \phi}{\partial x}\Big)^2 +\dfrac{\gamma_{\alpha\beta}}{2} \phi(1-\phi) \bigg)dx = \int_{-\infty}^{\infty} 
W^2_{\alpha\beta}\Big(\dfrac{\partial \phi}{\partial x}\Big)^2=\dfrac{\pi^2W^2_{\alpha\beta}}{8\eta},
\label{eq:sigma-ab}
\end{equation}
where we used Eqs.~(\ref{eq:static-equil-planar}) and (\ref{eq:sin-profile}) in the last steps.
Combining Eqs.~(\ref{eq:gamma-ab}) and (\ref{eq:sigma-ab}), the model parameters can be expressed in terms of surface free energy and interface thickness, as given above in Eqs.~(\ref{eq:gamma}) and (\ref{eq:W}).

\subsection{A single sphere}
\label{appSec:sphere}

For a single sphere of phase $\alpha$ in equilibrium with the surrounding medium, $\beta$, the radial symmetry of the problem can be used to write $\phi \equiv \phi(r)$, with $r$ being the distance from center of the sphere. As a consequence, one also can write $\nabla^2 \phi = \partial^2 \phi/\partial r^2 + ((d-1)/r)\partial \phi/\partial r$, where $d$ is dimensions of space. Substitution of this into Eq.~(\ref{eq:divP4}) and applying the equilibrium condition  yields,
\begin{equation}
(p_\alpha-p_\beta)\dfrac{\partial h}{\partial \phi} - \gamma_{\alpha\beta}\Big(\dfrac{1}{2}-\phi\Big) + W_{\alpha\beta}^2  \Big( \dfrac{\partial^2 \phi}{\partial r^2} + \dfrac{d-1}{r}\dfrac{\partial \phi}{\partial r}\Big)=0.
\label{eq:static-equil-sphere}
\end{equation}
Using Eq.~(\ref{eq:static-equil-sphere}), it is possible to obtain guidance regarding the choice of a reasonable interpolation function. For this purpose, we approximate $\phi(r)$ by the planar interface profile, Eq.~(\ref{eq:sin-profile}), (replacing, of course,  $x$ by $r-R$, with the sphere radius $R$). Within this planar approximation, we find $\partial \phi/\partial r =  -\pi/\eta\sqrt{\phi(1-\phi)}$ and $\partial^2 \phi/\partial r^2 =  (\pi/\eta)^2(1-2\phi)$. Inserting these expressions into Eq.~(\ref{eq:static-equil-sphere}) gives,
\begin{equation}
(p_\alpha-p_\beta)\dfrac{\partial h}{\partial \phi} =  \dfrac{8\sigma_{\alpha\beta}(d-1)}{\pi r} \sqrt{\phi(1-\phi)}.
\label{eq:dh-dphi1}
\end{equation}
Recalling that the product $\phi(1-\phi)$ is non zero only for values of $r$ in the interface region, one can replace $r$ by the  radius of sphere $R$. Further, using the Young-Laplace equation $p_\alpha-p_\beta = \sigma_{\alpha\beta} (d-1)/R$, Eq.~(\ref{eq:dh-dphi1}) simplifies to  

\begin{equation}
\frac{\partial h}{\partial \phi} = \dfrac{8}{\pi} \sqrt{\phi(1-\phi)}.
\label{eq:dh-dphi2}
\end{equation}
One can check that Eq.~(\ref{eq:dh-dphi2}) is satisfied by 
\begin{equation}
h(\phi) = \frac{1}{\pi}\Big(2(2\phi-1)\sqrt{\phi(1-\phi)} + \arcsin(2\phi-1) + \frac{\pi}{2} \Big).
\label{app:Interpolation_function3}
\end{equation}
\end{appendices}

\label{sect:bib}
\bibliographystyle{prsty_with_title}
\bibliography{references}
\end{document}